\documentclass[aps,prb,amsmath,amssymb,reprint,author-year,author-numerical,floatfix]{revtex4-2}
\usepackage{graphicx}
\graphicspath{{./images/}}
\usepackage{dcolumn}
\usepackage{bm}
\usepackage{mathtools}
\usepackage[caption=false]{subfig}
\usepackage{amssymb}
\expandafter\let\csname equation*\endcsname\relax
\expandafter\let\csname endequation*\endcsname\relax
\usepackage{amsmath}
\usepackage{amsbsy}
\usepackage{booktabs}
\usepackage{verbatim}
\usepackage{array}
\newcolumntype{P}[1]{>{\centering\arraybackslash}p{#1}}
\usepackage[colorlinks=true,citecolor=blue]{hyperref}
\AtBeginDocument{}%

\begin{document}

\preprint{AIP/123-QED}

\title{Theoretical Study of Electronic Transport in Two-Dimensional Transition Metal Dichalcogenides: Effects of the Dielectric Environment}
\author{Sanjay Gopalan}
\affiliation{Department of Materials Science and Engineering, The University of Texas at Dallas\\
             800 W. Campbell Rd., Richardson, TX 75080}
\author{Maarten L. Van de Put}
\altaffiliation[Present address: ]{imec, Kapeldreef 75, B-3001 Leuven, Belgium}
\affiliation{Department of Materials Science and Engineering, The University of Texas at Dallas\\
             800 W. Campbell Rd., Richardson, TX 75080}
\author{Gautam Gaddemane}
\altaffiliation[Present address: ]{imec, Kapeldreef 75, B-3001 Leuven,  Belgium} 
\affiliation{Department of Materials Science and Engineering, The University of Texas at Dallas\\
             800 W. Campbell Rd., Richardson, TX 75080}
\author{Massimo V. Fischetti}
\email[email: ]{max.fischetti@utdallas.edu.}
\affiliation{Department of Materials Science and Engineering, The University of Texas at Dallas\\
             800 W. Campbell Rd., Richardson, TX 75080}            

\date{\today}

\begin{abstract}
We discuss the effect of the dielectric environment (substrate/bottom oxide, gate insulator, and metal gates) on electronic transport in two-dimensional (2D)
transition metal dichalcogenides (TMD) monolayers. We employ well-known {\it{ab initio}} methods to calculate the low-field carrier mobility in free-standing
layers and use the dielectric continuum approximation to extend our study to layers in double-gate structures, including the effects of dielectric screening 
of the electron-phonon interaction caused by the bottom oxide and the gate insulator, and of scattering with hybrid interface optical-phonon/plasmon excitations
(`remote phonon scattering').
We find that the presence of insulators with a high dielectric constant may improve significantly the carrier mobility. However, scattering with the interface
hybrid excitations negates this gain and degrades the mobility significantly below its free-standing value. We find that this process is dominated by 
long-wavelength interactions that, for the carrier sheet-density of interest, are strongly affected by the coupling with the 2D plasmons.
Considering 2D layers in a double-gate geometry with SiO$_{2}$ as bottom-oxide and various top-gate insulators, we find that the mobility decreases as 
the top-insulator dielectric constant increases (from hBN to ZrO$_{2}$), as expected. However, we observe two main deviations from this trend: 
A high mobility is predicted in the case of the weakly polar hBN, and a mobility much lower than expected is calculated in the case of 
gate-insulator/TMD/bottom-oxide stacks in which two or more polar materials have optical-phonon with similar resonating frequencies. We also find that the 
effect of screening by metal gates is noticeable but not particularly strong. Finally, we discuss the effect of the TMD dielectric constant, of the free-carrier
density, and of temperature on the transport properties of TMD monolayers. 
\end{abstract}

\maketitle

\section{Introduction}

\subsection{Supported and gated 2D materials}
\label{sec:2Dmat}

The recent interest in two-dimensional (2D) materials, motivated by the search for alternatives to Si in the very large scale integration (VLSI) technology, has benefited from recent advances in density functional theory (DFT) that now permits the calculation not only of the excitation spectrum of a crystal, but also of
the electron-phonon matrix elements and of the low-field carrier mobility (see, for example, Refs.~\cite{Sohier_2018} and \cite{Ponce_2018}). Indeed,
limiting our attention to transition metal dichalcogenides (TMD) monolayers, that constitute the focus of our work,
{\it{ab initio}} methods have been used to predict the carrier mobility in a variety of them, for example by Jin and coworkers~\cite{Jin_2014}, 
by Gunst {\it{et al.}}~\cite{Gunst_2016}, by Sohier and coworkers~\cite{Sohier_2018}, and by Rawat {\it{et al.}}~\cite{Rawat_2018}. For MoS$_2$ in
particular, similar studies have been reported by Kaasjberg and coworkers~\cite{Kaasjberg_2012}, by Restrepo {\it{et al.}}~\cite{Restrepo_2014}, 
by Li and coworkers~\cite{Li_2015}, and by our own group~\cite{gaddemane2021limitations}. An {\it{ab initio}} study of quantum electron transport
in MoS$_2$ field effect transistors (FETs) has also been reported by Szab\'{o} {\it {et al.}}~\cite{Szabo_2015}.
 
Unfortunately, so far the vast majority of these first-principles studies have focused mainly on electronic transport in {\em {ideal free-standing}}
layers (with the notable exceptions discussed in the next subsection), 
ignoring the effect that the dielectric environment (that is, the insulating substrate, the gate insulator(s), and the metal gates) may have on the 
charge-transport characteristics of the material. These are more than mere deviations from ideality, since in almost all applications the monolayers are
supported by an insulating substrate and must be gated, implying the presence of a gate insulator and gate contact(s). Therefore, charge transport in these 
van der Waals (vdW) monolayers is controlled not only by their `bulk' properties, but also by the proximity of these polar insulators. Their influence goes 
beyond process-dependent effects, such as the presence of defects, charge impurities, non-ideal interfaces, which one may always hope 
to minimize by optimizing the technology, much as in the case of interface roughness for Si-based devices. Indeed, the dielectric polarization
of the top and bottom insulators may screen the `out-of-plane' electric-field lines, enhancing screening of all Coulomb interactions, including scattering
with the 2D phonons of the monolayers. Moreover, the plasma excitations of the two-dimensional electron gas (2DEG, the monolayer) and the optical 
phonons present in the system (substrate, gate insulator, and the 2D layer itself, if polar) contribute to the formation of hybrid interface excitations resulting from the coupling of these eigenmodes of the system. Scattering of electrons (and holes) with these hybrid interface plasmon/optical-phonon 
excitations (IPPs) is an additional process that affects electronic transport. This process goes under the somewhat improper name of `remote phonon scattering'. Here we shall avoid the use of this confusing term and refer to it more appropriately as to `IPP scattering'.

Before moving to the main subject of this work, we should point out that charge carriers may interact with excitations at the interface between the TMD layer and the
insulator not only via the polar (Fr\"{o}hlich-like) interactions that we consider here, but also via nonpolar (deformation-potential) interactions. 
An {\it{ab initio}} study of these nonpolar processes at the MoS$_2$/hBN interface has been presented recently by Fiore {\it{et. al}}~\cite{Fiore_2022}. 
Their effect on charge transport appears to be small, so we shall ignore them here. However, they may affect the thermal properties of material. 
\vspace*{-0.25cm}
\subsection{Overview of previous work on electron-IPP scattering}
\label{sec:previous} 

Scattering with these hybrid IPPs, or simply scattering with interface optical modes decoupled from 2D plasmons (SOs), has been investigated at length in the
past employing the dielectric continuum approximation (that we shall adopt here). Scattering of electrons with surface optical modes at the  Si/SiO$_{2}$ interface was first considered by Wang and Mahan~\cite{Wang_1972}, already in 1972, starting from Fuchs and Kliever's work~\cite{Fuchs_1966}. 
Taking hint from this study, the issue was reconsidered by Hess and Vogl~\cite{Hess_1979} (who coined the term 'remote phonons') -- ironically suggesting 
the use of low-$\kappa$ gate insulators to boost the electron mobility in Si inversion layers --, by Moore and Ferry~\cite{Moore_1980}, and 
by Bechstedt and Enderlein~\cite{Bechstedt_1985} -- accounting for the dependence of the process on the 2D carrier density -- in the 1980s.
Full coupling to polycrystalline-Si gate and inversion-layer 2D plasmons (important at the long wavelengths at which the interaction is strong) 
was considered by Fischetti {\it{et al.}}~\cite{Fischetti_2001} in 2001, although Toniutti {\it {et al.}}~\cite{Toniutti_2012},
in critically revisiting the problem a decade later, argued that plasmon-phonon hybridization does not play any major role (at least in the Si/SiO$_2$ 
system and at sufficiently short wavelengths. We shall show below that this is not the case, at least in the systems considered here). 

The study of electron scattering with the IPPs and/or SOs also been extended to other materials and interfaces: To supported and/or gated graphene~\cite{Fratini_2008,Konar_2010,Ong_2013,Wang_2013,Scharf_2013}, to carbon nanotubes~\cite{Perebeinos_2009}, to Si nanowires~\cite{Xiu_2011}, 
to III-V~\cite{Egdell_1987,Yu_1996,Hagen_2019,Ribeiro_2020}, and to III-nitrides heterostructures~\cite{Shi_2003,Singh_2019}. 

Coming to the structures considered here, several studies have been performed in the 
past using a variety of approximations: Zeng {\it{et al.}}~\cite{Lang_2013} considered supported and gated MoS$_2$ monolayers, ignoring coupling to 
plasmons. The dependence of the SO scattering strength on the dielectric environment in MoS$_2$ monolayers sandwiched 
between a substrate and a gate insulator was also studied by Ma and Jena~\cite{Ma_2014}, also ignoring the full coupling to plasmons and using a simplified
form of the scattering potential. The same simplified model was employed by Hosseini {\it {et al.}}~\cite{Hosseini_2015} to study strained TMDs.
Supported and gated InSe monolayers were considered by Chang and co-workers~\cite{Chang_2018}, also ignoring coupling to
plasmons, but considering the screening effect of a metal gate. 

Of particular interest is the work done by Hauber and Fahy ~\cite{Hauber_2017}. They have performed an analysis of the plasmon/optical-phonon excitations 
in several bulk III-V compound semiconductors and in supported-and-gated hBN and MoS$_2$ monolayers. 
Their analysis bypasses many of the approximations employed in all 
the previous studies that we have just mentioned. In particular, they have considered fully the plasmon-phonon coupling, also accounting (correctly) 
for the IPP lifetime (and, so, for Landau damping), treating electron and IPP transport self-consistently. 
However, given the complexity of their formulation, they have analyzed only one TMD system (supported and gated MoS$_2$ but without gate screening)
and a limited set of dielectrics (SiO$_2$ and HfO$_2$), in addition to vacuum. Moreover, they did not use {\it ab initio} methods to deal with
band-structure-related effects entering the calculation of other scattering processes (mainly, scattering with bulk TMD phonons). 

Finally, of note is a comprehensive study of IPPs in vdW heterostructures performed by Zhang and co-workers~\cite{Zhang_2019}, showing experimentally 
the existence of these modes and the validity of the long-wavelength approximation used to compute their dispersion. This is notable because the 
term 'remote phonons' has caused some confusion: As recently shown by Dyson and Ridley~\cite{Dyson_2020}, what may be properly called electron 
scattering with 'remote-phonons' (that is: electrons in the semiconducting layer interacting with the dipole field generated by optical phonons in an adjacent
insulator) is a process with negligible strength. Unfortunately, as we have already remarked, this term has come to label improperly the very different process 
we consider here, namely, electron scattering with the hybrid interface excitations. 

Experimental information that may confirm conclusively the importance of electron-IPP scattering is still controversial.
While the degradation of the mobility in channels supported and/or gated by high-$\kappa$ dielectrics is well established, alternative processes, 
such as remote Coulomb scattering~\cite{Saito_2002,Gamiz_2003,Barraud_2007,Esseni_2015} or scattering with interface dipoles~\cite{Ota_2007}, 
have been proposed to explain the observations. As a result, whereas several studies claim to have proven experimentally that the carrier mobility is 
controlled by electron-IPP scattering, (for example, Ref.~\cite{Chau_2004} for the Si/SiO$_2$ system, Ref.~\cite{Sonnet_2011} for the 
InGa$_{0.53}$As$_{0.47}$/HfO$_2$ stacks, or Ref.~\cite{Yu_2016} for MoS$_2$ supported by SiO$_2$, HfO$_2$, or Al$_2$O$_3$), other experimental studies 
have found that the observed mobility degradation is due mainly some of these alternative processes, such 
interface dipoles~\cite{Ota_2007} or Coulomb scattering~\cite{Tanimoto_2006}. 

Even more controversial is the possibility that an ideal metal gate may screen the IPP scattering potential: No evidence of this effect has been reported by 
Maitra {\it{et al.}} in Si metal-oxide-semiconductor FETs (MOSFETs)~\cite{Maitra_2007}; however, it has been observed in other
systems, such as organic and InGaZnO thin-film transistors~\cite{Ma_2019,Su_2021}. Laikhtman and Solomon have even predicted theoretically anti-screening 
effects due to a metal gate~\cite{laikhtman2008remote}. 

Another gate-related effect is the predicted depression of the electron mobility in polycrystalline Si (poly-Si) 
gate MOSFETs due to a resonance between gate plasmons and insulator optical-phonons. This has been proposed by 
Kotlyar {\it{et a.}}~\cite{kotlyar2004inversion}, by Shah and De~Souza~\cite{Shah_2007}, and by Suleiman {\it{et al.}}~\cite{suleiman2012effects}. This, as well
as the other issue mentioned above, will be discussed below. 

\subsection{Aim and organization of this work}
\label{sec:thiswork}

The aim of this work is to provide a comprehensive study of the effect of IPP scattering on the low-field mobility of electrons and holes, mostly at
300~K, employing physical models -- within the dielectric continuum approximation -- that bypass several approximations that are often used. Namely,
we employ {\it{ab initio}} (DFT) methods to calculate the band structure of the TMDs, the 2D 'bulk' phonon dispersion, and the electron-phonon matrix elements;
we account fully for the presence of the dielectrics by considering not only scattering with fully hybridized IPPs but also for for the screening effects of
the insulators on the strength of the interaction with bulk 2D phonons (an effect not considered previously) and, finally, we consider also the role played 
by nearby ideal-metal gates in screening both the bulk electron-phonon and IPP scattering. Finally, we use full-band Monte Carlo simulations to calculate 
the low-field mobility, thus addressing concerns raised by Toniutti {\it{et al.}}~\cite{Toniutti_2012}. In this study we consider a variety of 
hexagonal TMD monolayers (MoS$_2$, MoSe$_2$, MoTe$_2$, WS$_2$, WSe$_2$, and also WTe$_2$\cite{bnote3}) in a double-gate structure `sandwiched' between 
a SiO$_2$ bottom-gate-insulator and several gate insulators (SiO$_2$, hBN, AlN, Al$_2$O$_3$, HfO$_2$, and ZrO$_2$). 

The main results of our work can be summarized as follows: The presence of a bottom insulator and/or a gate insulator with a high dielectric constant 
(high-$\kappa$ insulators) screens the interaction of electrons (and holes) with the bulk (2D) phonons of the monolayer. Whereas, in principle, this has a strong 
beneficial effect on the carrier mobility, IPP scattering depresses the mobility below its free-standing value, with the notable exception of hBN, 
as already found by Ma and Jena~\cite{Ma_2014}, since it is only
weakly polar and the frequencies of its optical phonons are large enough to be decoupled from all other frequencies entering the problem. Moreover, the carrier
mobility follows the expected trend, decreasing with increasing dielectric constant of the gate insulator; however, there are significant deviations from this behavior:
In gate-insulator/TMD/SiO$_2$ structures in which two or more optical-phonon frequencies are similar, resonance of these modes causes an enhanced 
interface polarization charge, resulting in a very low mobility. Finally, the screening effect of the ideal-metal gate is noticeable, but not particularly strong. 

We organize our discussion as follows:
In Section~\ref{sec:TM}, we present the theoretical model and computational methods used to calculate the carrier mobility in different monolayer TMDs 
accounting for presence of different dielectric environments. In Section~\ref{sec:results}, we present results for the carrier mobility in the various
TMD monolayers we have considered and discuss in-detail how the dielectric environment affect their transport properties. Finally, we present our conclusions in Sec.~\ref{sec:Conclusions}. 

\section{Theoretical Model}
\label{sec:TM}

Throughout our discussion, we use the dielectric-continuum limit, the long-wavelength limit of the dielectric ionic response of all materials; and, for the 
insulators, we ignore details about their morphology (crystalline, polycrystalline, or amorphous), treating them as isotropic amorphous media.
Furthermore, we consider only a maximum of two optical phonons per material, those with the largest oscillator strength, 
as measured by Fourier-transform infrared spectroscopy, as explained in Ref.~\cite{Fischetti_2001}.  

As a matter of notation, we employ lower-case bold symbols (such as $\mathbf{r}$ or $\mathbf{k}$) to denote 3-vectors and
upper-case bold symbols (such as $\mathbf{R}$ or $\mathbf{K}$) to denote 2-vectors on the $(x,y)$ plane of the layer. We use SI units and
denote by $e$ the magnitude of the electron charge, by $m_{0}$ the free electron mass, by $\epsilon_{0}$ the permittivity of vacuum, by $\hbar$ the reduced
Planck's constant, by $T$ the absolute temperature, and by $k_{\rm B}$ Boltzmann's constant. Other quantities will be defined as needed. Finally, to avoid a cumbersome notation with too many subscripts, occasionally we shall be guilty of using an improper notation by denoting the functional dependence of 
several quantities on independent variables as subscripts or within parentheses ({\it {e.g.}}, $\epsilon(Q,\omega)$ or $\epsilon_{Q,\omega}$), regardless of
whether we use finite-volume or infinite-volume normalization.   
\begin{figure}[tb]
\centerline{\includegraphics[width=8.750cm]{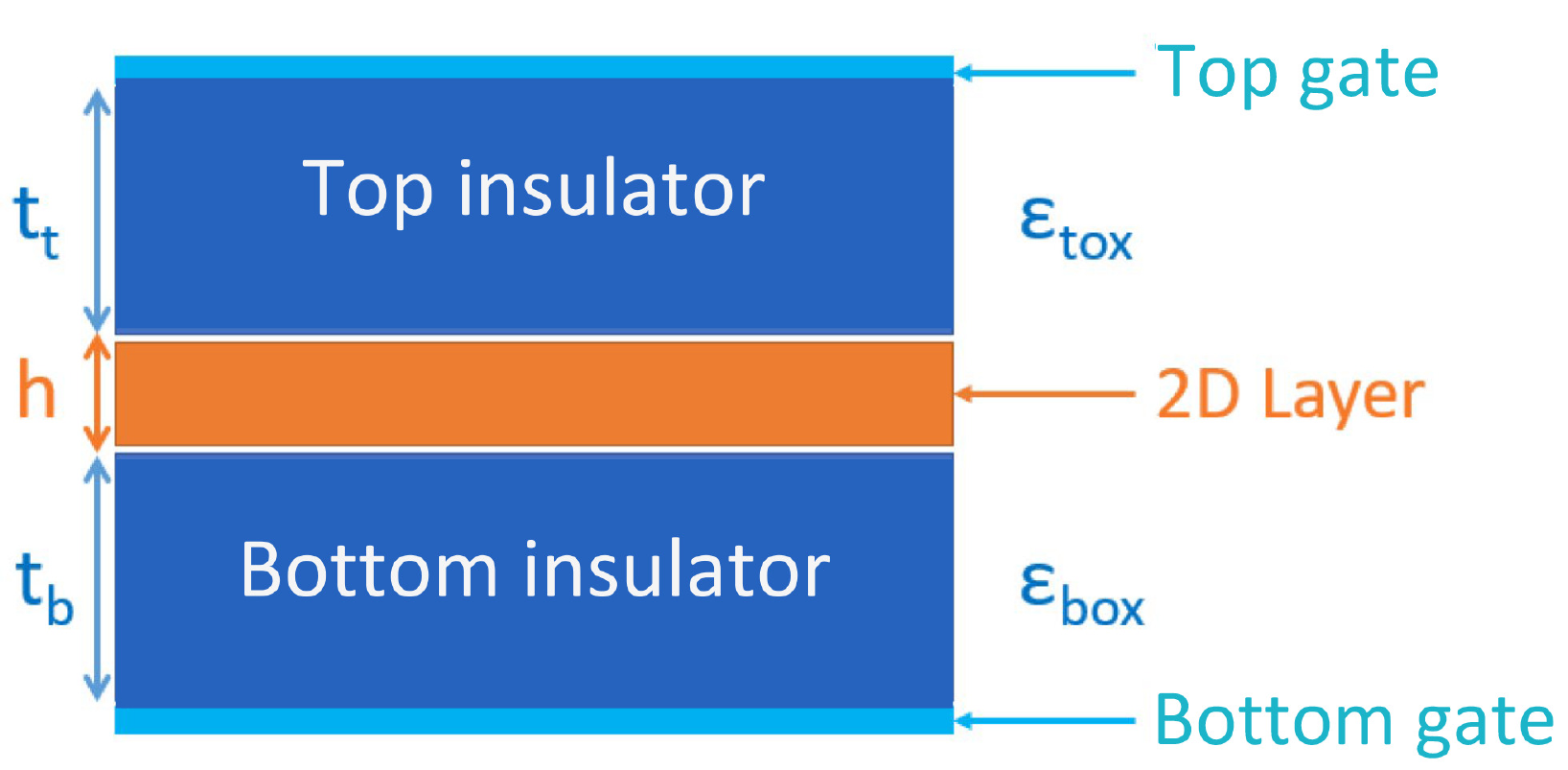}}
\caption{Schematic diagram of the double-gated TMD monolayer considered here.}
\label{fig:System}
\end{figure}

\subsection{The system}
\label{sec:system}

The system we consider, shown in Fig.~\ref{fig:System}, consists of an ideal (bottom) metal gate in the region $z \le -t_{\rm b}$; an insulator (which can be 
viewed either as a substrate, when $t_{\rm b} \rightarrow \infty$), or as a bottom gate insulator) with dielectric function $\epsilon_{\rm box}(\omega)$ in the
region $-t_{\rm b} < z \le 0$; a 2D layer of thickness $h$ and dielectric function $\epsilon_{\rm 2D}(Q,\omega)$ in the region $0 < z \le h$; a gate insulator with
dielectric function $\epsilon_{\rm tox}(\omega)$ in the region $h < z \le h+t_{\rm t}$; an ideal metal in the half-space $z > h+t_{\rm t}$. Ignoring the confinement of the lattice vibrations in thin layers, an effect that may weaken the ionic dielectric response of thin dielectrics and has been studied by
Laikhtman and Solomon~\cite{laikhtman2008remote}, for the insulators we assume the 'bulk' long-wavelength limit of the dielectric functions, 
dependent on the frequency $\omega$, assuming only two optical phonons in each oxide, as explained above:
\vspace*{-0.25cm}
\begin{multline}
\epsilon_{\rm box}(\omega ) \ =  \ \epsilon_{\rm box}^{(\infty)} + 
                     [ \epsilon_{\rm box}^{(0)} - \epsilon_{\rm box}^{\rm (mid)} ] 
                         \frac{\omega_{\rm TO,1}^{2}}{\omega_{\rm TO,1}^{2} - \omega^{2}} + \\  
                     [ \epsilon_{\rm box}^{\rm (mid)} - \epsilon_{\rm box}^{(\infty)} ] 
                         \frac{\omega_{\rm TO,2}^{2}}{\omega_{\rm TO,2}^{2} - \omega^{2}} \ = 
                    \epsilon_{\rm box}^{(\infty)} \frac{ \omega_{\rm LO,1}^{2}-\omega^{2} }{\omega_{\rm TO,1}^{2}-\omega^{2} }
                                                   \frac{ \omega_{\rm LO,2}^{2}-\omega^{2} }{\omega_{\rm TO,2}^{2}-\omega^{2} } \ ,
\label{eq:epsbox} 
\end{multline}
\vspace*{-0.25cm}
\begin{multline}
\epsilon_{\rm tox}(\omega ) \ =  \ \epsilon_{\rm tox}^{(\infty)} + 
                     [ \epsilon_{\rm tox}^{(0)} - \epsilon_{\rm tox}^{\rm (mid)} ] 
                         \frac{\omega_{\rm TO,3}^{2}}{\omega_{\rm TO,3}^{2} - \omega^{2}} + \\ 
                     [ \epsilon_{\rm tox}^{\rm (mid)} - \epsilon_{\rm tox}^{(\infty)} ] 
                         \frac{\omega_{\rm TO,4}^{2}}{\omega_{\rm TO,4}^{2} - \omega^{2}} \ = 
                    \epsilon_{\rm tox}^{(\infty)} \frac{ \omega_{\rm LO,3}^{2}-\omega^{2} }{\omega_{\rm TO,3}^{2}-\omega^{2} }
                                                   \frac{ \omega_{\rm LO,4}^{2}-\omega^{2} }{\omega_{\rm TO,4}^{2}-\omega^{2} } \ ,
\label{eq:epstox} 
\end{multline} 
\noindent having indicated with $\omega_{\rm TO,1}$ and $\omega_{\rm TO,2}$ the low- and high-frequency optical phonons of the substrate insulator,
with $\omega_{\rm TO,3}$ and $\omega_{\rm TO,4}$ the low- and high-frequency optical phonons of the gate insulator, and with 
$\epsilon^{(0)}$, $\epsilon^{\rm (mid)}$ and $\epsilon^{(\infty)}$ the static, intermediate, and optical dielectric constants of the insulators. 
As mentioned above, since we assume amorphous insulators, it is reasonable to treat them as isotropic dielectrics, 
we do not need to specify whether these dielectric functions are transverse or longitudinal components of the full dielectric tensor. This is not 
the case for the 2D TMD monolayer. In this case, we assume for the out-of-plane dielectric function for the 2D layer:
\begin{multline}
\epsilon_{\rm 2D \perp}(Q,\omega) = \epsilon_{\rm 2D \perp}^{(\infty)} \left [ 1 - e^{2} G_{Q}(h/2,h/2) \Pi_{\rm 2D}(Q,\omega) \right ] \ + \\ 
    [ \epsilon_{\rm 2D \perp}^{(0)} - \epsilon_{\rm 2D \perp}^{(\infty)} ] \frac{\omega_{\rm ZO}^{2}}{\omega_{\rm ZO}^{2}-\omega^{2}}  \ ,
\label{eq:eps2D}
\end{multline} 
\noindent where $Q$ is the magnitude of the wave vector on the plane of the layer,
$\epsilon_{\rm 2D \perp}^{(\infty)}$ and $\epsilon_{\rm 2D \perp}^{(0)}$ are the the out-of-plane optical and static dielectric constants 
(as calculated via DFT together with the dielectric thickness $h$, as given in Ref.~\cite{laturia2018dielectric}, or via the formalism 
of Ref.~\cite{tian2019electronic}). The function $G_{Q}(z,z')$ is the in-plane Fourier transform of the Poisson Green’s function for the geometry of interest; 
it represents the potential at
$z=h/2$ (in the middle of the monolayer) caused by a point-charge also at $z'=h/2$. This will be discussed below (see Eqs.~(\ref{eq:Genv})-(\ref{eq:lambda2}))
for the general double-gate geometry. The quantity $\Pi_{\rm 2D}(Q,\omega)$ is the polarizability of the free carriers in the 2D layer as given, for
example, by Stern~\cite{Stern_1967} or by the simpler approximation given below. Finally, the last term in Eq.~(\ref{eq:eps2D}) reflects the out-of-plane 
ionic polarization of the monolayer; obviously, it is present only in the case of a polar 2D layer with out-of-plane optical phonons, ZO, with frequency
$\omega_{\rm ZO}$~\cite{Zhang_2020}. 

Regarding the anisotropy of the dielectric constant of the TMD layer, an approximation that bypasses several subtle issues, and that we employ here,
consists in defining an `effective' dielectric constant 
$\epsilon_{\rm 2D}^{\rm (eff)} =  \epsilon_{\rm 2D \perp} (\epsilon_{\rm 2D \parallel}/\epsilon_{\rm 2D \perp})^{1/2}$, as suggested by 
Hauber and Fahy~\cite{Hauber_2017}. Regardless of the assumption made, in the following we shall omit the subscripts `$\perp$' or superscript `eff', 
since only the out-of-plane or effective dielectric function of the 2D layer will be used. 

We must stress that in the $\sigma_{\rm h}$-symmetric layers we consider here ({\it {i.e.}}, the 2H TMD monolayers that are symmetric under reflections 
on the $(x,y)$ plane), the polarization caused by the ZO-phonon polarization induces a potential that is antisymmetric with respect to the layer of the 
transition-metal ions, Mo or W~\cite{Zhang_2020}. Therefore, at first-order, the matrix elements of the scattering potential associated with such an 
antisymmetric potential vanishes. In other words, electrons do not couple to the out-of-plane (ZO) component of the hybrid modes that we shall consider below.   

At a finite temperature, the expression for the zero-temperature electronic polarizability $\Pi_{\rm 2D}(Q,\omega)$ given 
by Stern~\cite{Stern_1967}, could be replaced by the well-known finite-temperature expression given by Maldague~\cite{Maldague_1978}. 
Unfortunately, its use results in numerical complications, since solving the secular equations that we shall consider below -- required to obtain the dispersion of the hybrid interface modes of interest here -- would not consist on finding the roots of a polynomial, but on the
much more difficult task of finding the roots of a transcendental equation.

This difficulty can be circumvented by noticing that the expressions given by Eqs.~(\ref{eq:epsbox}) and (\ref{eq:epstox}) are valid only in the 
long-wavelength limit $Q \rightarrow 0$. Therefore, in the same long-wavelength spirit, the electronic contribution to Eq.~(\ref{eq:eps2D}) (that is,
of the free-carriers term given by the first term at the right-hand side of Eq.~(\ref{eq:eps2D})) can be approximated by the simpler expression:
\begin{equation}
\epsilon_{\rm 2D}^{\rm (el)}(Q,\omega) \approx \epsilon_{\rm 2D}^{(\infty)} \left [ 1 - \frac{\omega_{\rm P}(Q)^{2}}{\omega^{2}} \right ] \ ,
\label{eq:eps2D1}
\end{equation} 
having defined the 2D plasma dispersion $\omega_{\rm P}(Q)^{2} = e^2 n Q/(2 \epsilon_{2D}^{\infty} m^{\ast})$, where $n$ is the density 
of the free carriers and $m^{\ast}$ their effective mass.
This expression captures the small-$Q$ behavior in the region in which the response is not strongly damped by a large imaginary part; that is,
outside the single-particle region in which $E({\bf k}_{\rm F} + {\bf Q}) - E({\bf k}_{\rm F}) = \hbar \omega$ (${\bf k}_{\rm F}$ being the Fermi wave vector).
Moreover, in this range of long wavelengths, there is no significant dependence on temperature between 0 and 300~K. Therefore, to simplify the numerical
complexity of the formulation and without making any severe error, we can use Eq.~(\ref{eq:eps2D1}) as long as wavelength of the perturbation is longer
than the `Landau damping' wavelength, $\sim 1/Q_{\rm LD}$, defined implicitly by:
\begin{equation}
Q_{\rm LD} \ = \ \left [ k_{\rm F}^{2} + \frac{ 2 m^{\ast} \omega_{\rm P}(Q_{\rm LD}) }{ \hbar } \right ]^{1/2} - k_{\rm F} \ . 
\label{eq:LD2}
\end{equation}
The use of this expression in the long-wavelength limit, $Q \rightarrow$ 0, is also consistent with the range of validity of Eqs.~(\ref{eq:epsbox}) and
(\ref{eq:epstox}). 
On the contrary, for $Q > Q_{\rm LD})$, we may treat the response of the free carriers in the 2D layer as purely static using the $Q$-dependent Thomas-Fermi expression, as discussed below before Eq.~(\ref{eq:QTFnd}). The problem of how to account correctly for Landau damping -- and how we have tackled the problem --
will be discussed below in Sec.~\ref{sec:Landau}.

We have calculated the plasma frequency, $\omega_{\rm P}(Q)$ that appears in Eq.~(\ref{eq:eps2D1}) ignoring the possibility that electrons populating the
satellite Q valleys (sixfold degenerate minima located along the K–$\Gamma$-direction) may result in a different plasma response. 
The coupled plasma oscillations of two two-dimensional electron gases, characterized by a different 
effective mass and density, have been studied by Vignale~\cite{Vignale_1987}. He found that this situation results in the presence of two plasma modes: 
A `fast' optical plasmon corresponding, in our case, to  electrons in the K and Q valleys oscillating in phase, and a `slow' acoustic mode associated with 
K and Q electrons oscillating out of phase. However, in all the materials and in the near-equilibrium situations considered here, the energy of the satellite 
Q-valleys is sufficiently larger than $k_{\rm B}T$ as to render the electron density in this valleys only a small fraction of the total density. For example,
the six satellite Q-valleys in MoS$_{2}$ are at an energy between 70 and 270~meV (the value we obtain using the pseudopotentials and exchange-correlation functionals chosen here~\cite{gaddemane2021limitations}) above the minima at the symmetry points K. Therefore, we have simplified the problem 
by employing Eq.~(\ref{eq:eps2D1}) to account for the response of the optical plasmon, while ignoring the effect of the few electrons in the satellite valleys. 
Moreover, we have ignored the acoustic plasmon since, as shown by Vignale~\cite{Vignale_1987}, for a small occupation of the satellite valleys, 
it will be strongly Landau-damped by the fast optical plasmon.  

\begin{table}
\caption{Parameters used for the density functional theory (DFT) and density functional perturbation theory (DFPT) calculations.}
\vspace*{0.5cm}
\begin{tabular}{ll}
\hline
\hline
Parameters                       & \\
\hline
Kinetic energy (E$_{k}$) cutoff  & 60 Ry  \\
Charge density cutoff            & 240 Ry\\
Ionic minimization threshold     & 10$^{-6}$ Ry\\
Self-consistent field threshold  & 10$^{-12}$ Ry\\
$\mathbf{k}$-point mesh          & 12$\times$12$\times$1\\
\hline
\hline
\end{tabular}
\label{tab:dftpara}
\end{table}

\subsection{{\emph{Ab initio}} calculations}
\label{sec:abinitio}

In order to obtain the electronic band structures, phonon dispersion, and electron-phonon matrix elements in the 2H-TMDs of interest, we have
employed {\it{ab initio}} methods that have become almost `routine'. Here, we shall give here only the essential information. We should note that, in obtaining the 
results presented in this section, we have ignored possible corrections caused by a non-zero temperature and the presence of a gate bias.
   
We have used density functional theory (DFT) as implemented in the Quantum ESPRESSO (QE) package~\cite{giannozzi2009quantum} with the 
Perdew–Burke–Ernzerhof generalized-gradient approximation (GGA-PBE)~\cite{perdew1996generalized}
for the exchange-correlation functional, and the norm-conserving Vanderbilt pseudopotentials (ONCV)~\cite{hamann2013optimized} for each constituent
element. The atomic structure for these materials is found by minimizing the total energy with respect to the lattice constants and ionic positions. 
The computational parameters used in these calculations are shown in Table~\ref{tab:dftpara}. As required by the Monte Carlo simulations described in
Ref.~\cite{gaddemane2018theoretical,gaddemane2021limitations,gaddemane2021monte} and used here, the band structure is tabulated on a fine
mesh, 201×201×1, covering a rectangular section that inscribes the triangular irreducible wedge of the hexagonal first Brillouin zone.
For the calculation of the phonon dispersion and of the electron–phonon matrix elements, we have used the `Electron–Phonon Wannier' (EPW) software package~\cite{giustino2007electron,baroni2001phonons,giustino_2019,giustino2019erratum} which uses density functional perturbation theory (DFPT). 
These quantities are initially calculated on a coarse $\mathbf{k}$- (12×12×1) and $\mathbf{q}$-mesh (6×6×1) and interpolated on finer 
$\mathbf{k-}$ (30×30×1) and $\mathbf{q}$-meshes (30×30×1) using maximally-localized Wannier functions.
The coarse and the fine meshes span the entire Brillouin zone. The phonon dispersion and electron–phonon matrix elements on the fine mesh are finally
interpolated on the band structure mesh using a bilinear interpolation. The phonon-limited electron–phonon scattering rates are calculated using Fermi’s
golden rule, and tabulated on the same mesh of $\mathbf{k}$-points used to tabulate the band structure.

We have not considered the effect of spin-orbit coupling (SOC) since, as shown in our previous work~\cite{gaddemane2021limitations}, it affects the calculated 
electron mobility only marginally, by less than 10\%, in the cases of MoS$_2$ and MoTe$_2$, although its effect is larger for WS$_2$ and it may affect the hole
mobility more significantly. As we shall see below, the effects of dielectric screening of the surrounding insulators and of IPP scattering
are much larger then those caused by SOC. Therefore, we expect that ignoring the numerically cumbersome effects of SOC is not going to alter 
significantly the qualitative trends of our results and, in the cases of  MoS$_2$ and MoTe$_2$, not even quantitatively.     

\subsection{Screening of the electron-phonon interaction by the dielectric environment}
\label{sec:screen}

The electron-phonon matrix elements calculated using DFT assume free-standing layers, since the calculations are performed using a supercell consisting of 
the isolated layer surrounded by sufficiently 'thick' vacuum padding to isolate adjacent supercells. Of course, it would be desirable to perform the 
calculations using a more complicated and larger supercell that includes the dielectrics and the gates, also accounting for the presence of free carriers 
in the monolayer. In so doing, one would obtain information about confined phonons in the insulators and interface polarization charges, for example. 
However, this task is extremely complex. 
Therefore, in order to consider how the electron-phonon interaction is affected by the presence of the dielectrics and of the metal gates, 
we consider how the electron-phonon matrix elements calculated using DFT depend on the Poisson Green's function and replace it with the Green's
function that satisfies the boundary conditions of the double-gate structures of interest here. Although we are forced to make use of some 
simplifications and approximations, the procedure should capture the main physics of the process. 

First, we note that the matrix elements between two Bloch states ${\bf k}$ in band $n$ and ${\bf k}^{\prime}$ in band $n^{\prime}$ obtained 
from DFT have the form:
\begin{equation}
\label{eqn:el-ph}
    \langle \bf{k}^{\prime},n^{\prime} | \delta E^{(\eta)}_{\rm tot} | \bf{k},n \rangle = 
       \int {\rm d} \bf{r} \ \psi^{*}_{n^{\prime}, \bf{k}^{\prime}}(\bf{r}) \  
            \delta E^{(\eta)}_{\rm tot}(\bf{r}) \ \psi_{n,\mathbf{k}} (\mathbf{r}) \ ,
\end{equation}
\noindent where $\psi_{n,{\bf{k}}}({\bf{r}}) = e^{i {\bf k} \cdot {\bf r}} \sum_{\bf {g}} u^{(n)}_{{\bf k},{\bf g}} e^{i {\bf g} \cdot {\bf r}}/\Omega^{1/2}$
is the Bloch wave of the TMD layer ($\Omega=N_{\rm cell} \Omega_{\rm cell}$ is the normalization volume and the 3-vectors ${\bf g}$ are the vectors in 
reciprocal space) and $\delta E^{(\eta)}_{\rm tot}(\mathbf{r})$ is the change of the total
energy of the system caused by a phonon of branch (acoustic/optical, longitudinal/transverse/flexural) $\eta$.
Perhaps too pedantically, we have expressed the electronic states as dependent on the 3-vector ${\bf k}=({\bf K},k_{z})$, since they are obtained 
from DFT calculations 
that that employ a three-dimensional supercell. However, since we are interested only in states with energy much below the vacuum level, we could 
ignore their dependence on the out-of-plane component, $k_{z}$, and write, instead, $\psi_{\mathbf{K}}(\mathbf{r})$, a notation that we shall use in the following.  

Both the Hartree and the 
exchange-correlation potentials contribute to this change; that is (oversimplifying the notation in writing exchange and correlation as a local potential), 
$\delta E^{(\eta)}_{tot}(\mathbf{r}) = \delta E^{(\eta)}_{\rm H}(\mathbf{r})+\delta E^{(\eta)}_{\rm xc}(\mathbf{r})$. These are the quantities that need to be 
modified when considering the more complicated geometry of interest. Now we note that the Hartree energy,
$E_{\rm H}(\mathbf{r})$, depends linearly on the electron density, whereas $E_{\rm xc}(\mathbf{r})$ has a weaker dependence 
(namely, $E_{\rm xc}(\mathbf{r}) \sim \rho^{(\eta)}(\mathbf{r})^{1/3}$ in the local-density approximation). Therefore, we expect that the 
exchange-correlation terms will have a contribution to the scattering rates smaller than the Hartree term under a modification of the dielectric environment.
Thus, we consider only the Hartree component whose contribution, in the case of a free-standing layer, as done using DFT, can be written as:
\vspace*{-0.2cm}
\begin{equation}
\label{eqn:hartree}
    \delta E_{\rm tot}(\mathbf{r}) \; \approx \; \delta E^{(\eta)}_{\rm H}(\mathbf{r}) \ = \ \int d\mathbf{r}^{\prime} \ 
                                G^{(0)}(\mathbf{r},\mathbf{r}^{\prime}) \; \delta\rho^{(\eta)}(\mathbf{r}^{\prime}) \ ,
 \vspace*{0.2cm}
\end{equation}
\noindent where $G^{(0)}(\mathbf{r},\mathbf{r}^{\prime})$ is the Green's function for the Poisson equation {\it {in vacuo}} and
$\delta\rho^{(\eta)}(\mathbf{r}^{\prime})$ is the change of the charge at $\mathbf{r}^{\prime}$ induced by a phonon of branch $\eta$.
This can be written as:
\begin{equation}
\delta \rho^{(\eta)}({\bf r}') \ = \ 
      \sum_{\alpha l} \ \widetilde{\nabla}_{{\bf R}_{\alpha l}} \rho ({\bf r}') \cdot \delta {\bf r}^{(\eta)}_{\alpha l} \ ,
\label{eq:deltarho}
\end{equation}
where the quantity $\delta {\bf r}^{(\eta)}_{\alpha l}$ is the displacement of the ion $\alpha$ in cell $l$ due to phonons of branch $\eta$.
The symbol $\widetilde{\nabla}$ should be interpreted as a `functional gradient', so that the function 
$\widetilde{\nabla}_{{\bf r}_{\alpha l}} \rho({\bf r})$ (often written as $\delta \rho({\bf r})/\delta{\bf r}_{\alpha l}$)
represents the change of the electronic charge $\rho$ at position ${\bf r}$ under an infinitesimal shift of ion $\alpha$ in cell $l$ 
along the Cartesian directions. 
This can be obtained from the ‘small displacement method’ (implemented, for example, in the computer program PHON~\cite{Alfe_2009}) or 
from DFPT~\cite{baroni2001phonons}. 
We now write ${\bf r}_{\alpha l} = {\bf r}_{l} + \boldsymbol{\tau}_{\alpha}$, where ${\bf r}_{l}$ is the lattice site
$l$ and $\boldsymbol{\tau}_{\alpha}$ is the position of ion $\alpha$ in the unit cell. 

We now consider only in-plane phonons (since, as explained above, scattering with out-of-plane vibrations is forbidden at first order in the symmetric 
TMDs we consider), also consider negligible their displacement and any change of the charge density they induce along the $z$ direction. 
Therefore, using the usual expression for 
$\delta {\bf r}^{(\eta)}_{\alpha l}$ and tracing over the phonon states assumed to be at thermal equilibrium, the charge density obtained from DTT,
 Eq.~(\ref{eq:deltarho}), can be expressed approximately as: 
\begin{multline}
\delta \rho^{(\eta)}({\bf r}') \ \approx \ 
      \sum_{\alpha l {\bf Q}} \ \widetilde{\nabla}_{{\bf r}_{\alpha l}} \rho ({\bf r}') \ e^{i{\bf Q} \cdot {\bf r}_{\alpha l}} 
        \cdot {\bf e_{\parallel}}^{(\eta, \alpha)}_{\bf Q} \\
         \times \left ( \frac{\hbar}{2 M_{\alpha} \omega^{(\eta)}_{\bf Q}} \right )^{1/2} 
            \left \{ \begin{array}{c} N(\omega^{(\eta)}_{\bf Q})^{1/2} \\ \left [ 1+N(\omega^{(\eta)}_{\bf Q})\right ]^{1/2} \end{array} \right \} \ = \\
       \sum_{\alpha l {\bf Q}} \ A^{(\eta, \alpha )}_{\bf Q} \ e^{i{\bf Q} \cdot {\bf r}_{\alpha l}} \
               {\bf e_{\parallel}}^{(\eta, \alpha)}_{\bf Q} \cdot \widetilde{\nabla}_{{\bf r}_{\alpha l}} \rho ({\bf r}')  \ ,
\label{eq:deltaR}
\end{multline} 
where the quantity $A^{(\eta, \alpha )}_{\bf Q}$ is defined implicitly by the last equality, 
${\bf e_{\parallel}}^{(\eta, \alpha)}_{\bf Q}$ and $\omega^{(\eta)}_{\bf Q}$ are the in-plane polarization (for the ion $\alpha$) and frequency of a phonon of branch $\eta$ and wave vector ${\bf Q}$, $N(\omega^{(\eta)}_{\bf Q})$ is the Bose-Einstein
occupation of such phonons, and $M_{\alpha}$ is the mass of the ion $\alpha$. The upper and lower quantities in the curly bracket refer to absorption and
emission processes, respectively. Also, note that ${\bf r}_{\alpha l}$ is a 3-vector but ${\bf Q}$ is a 2-vector; thus, the inner product
${\bf Q} \cdot {\bf r}_{\alpha l}$ should be understood as ${\bf q} \cdot {\bf r}_{\alpha l}$, where ${\bf q} = ({\bf Q},0)$. Similarly, 
${\bf e_{\parallel}}^{(\eta, \alpha)}_{\bf Q} \cdot \widetilde{\nabla}_{{\bf r}_{\alpha l}} \rho ({\bf r}')$ denotes the dot-product of the in-plane
components of these vectors. In the following, we shall denote by ${\bf R}_{\alpha l}$ and ${\bf R}_{l}$ the in-plane components of 
${\bf r}_{\alpha l}$ and ${\bf r}_{l}$, respectively. 

Using this expression and the Bloch form of the wavefunctions, the Hartree-only component of the DFT matrix element given by Eq.~(\ref{eqn:el-ph}) 
is approximately:
\begin{widetext}
\begin{multline}
\langle {\bf K'},n^{\prime}  | \delta E^{(\eta)}_{\rm H} | {\bf K},n \rangle  \approx 
   \sum_{{\bf g}' {\bf g}} \ u^{(n') \ast}_{{\bf K}', {\bf g}'} \ u^{(n)}_{{\bf K}, {\bf g}}  \ \frac{1}{\Omega} \
        \int {\rm d}{\bf R} \ e^{i ({\bf K}'-{\bf K}+{\bf G}'-{\bf G}+{\bf Q}') \cdot {\bf R}} \
        \int {\rm d} z \ e^{i(g_{z}'-g_{z})z}  \\
       \times \sum_{\alpha l {\bf Q} {\bf Q}'} \ A^{(\eta, \alpha)}_{\bf Q} \ 
             e^{i({\bf Q}-{\bf Q}') \cdot {\bf R}_{l}} \ e^{i({\bf Q}-{\bf Q}') \cdot \boldsymbol{\tau}_{\alpha}} \
               {\bf e}^{(\eta, \alpha)}_{\parallel,{\bf Q}} \cdot \int {\rm d}{\bf R}' 
                 e^{-i{\bf Q}' \cdot ({\bf R}'-{\bf R}_{\alpha l})} 
                  \int {\rm d} z' G^{(0)}_{{\bf Q}'}(z,z') \ \widetilde{\nabla}_{{\bf R}_{\alpha l}} \rho ({\bf R}',z') \ ,
\label{eq:matrixel4}
\end{multline}
having expressed $G^{(0)}({\bf r},{\bf r}')$ in terms of its in-plane Fourier components.
\end{widetext}
In this expression, the factor
\begin{multline}
\sum_{\alpha} \ \ e^{i({\bf Q}-{\bf Q}') \cdot \boldsymbol{\tau}_{\alpha}} \
        \int {\rm d}{\bf R}' e^{-i{\bf Q}' \cdot ({\bf R}'-{\bf R}_{\alpha l})} \\ \times
                  \int {\rm d} z' G^{(0)}_{{\bf Q}'}(z,z') \widetilde{\nabla}_{{\bf R}_{\alpha l}} \rho ({\bf R}',z') 
\label{eq:deltarho2}
\end{multline}
does not depend on the cell index $l$ thanks to the periodicity of the 2D crystal, 
so the factor $\sum_{l} e^{i({\bf Q}-{\bf Q}') \cdot {\bf R}_{l}}$ appearing in Eq.~(\ref{eq:matrixel4}) is nonzero only
when ${\bf Q}={\bf Q}'$, up to a vector of the reciprocal lattice (which ultimately will be ignored).

Since dielectric screening -- whose effects we are considering here -- is particularly strong at long wavelengths, much longer than the thickness of the TMD monolayer, we can consider  
the change of the charge $\widetilde{\nabla}_{{\bf R}_{\alpha l}} \rho ({\bf R}',z')$
as localized at $z'=d$ (so, is proportional to $\delta(z'-d)$). Thus, the term given by Eq.~(\ref{eq:deltarho2}) can be approximated by:
\begin{multline}
\hspace*{-0.25cm}
\sum_{\alpha} \ e^{- i {\bf Q}' \cdot \boldsymbol{\tau}_{\alpha}} \ 
      \int {\rm d}{\bf R}' e^{i{\bf Q}' \cdot ({\bf R}'-{\bf R}_{l})} \
             G^{(0)}_{{\bf Q}'}(z,d) \ \widetilde{\nabla}_{{\bf R}_{\alpha l}} \rho ({\bf R}',d) \\ = \
                 \ G^{(0)}_{{\bf Q}'}(z,d) \ \sum_{\alpha} \ 
                    e^{- i {\bf Q}' \cdot \boldsymbol{\tau}_{\alpha}} \ [\widetilde{\nabla}_{\boldsymbol{\tau}_{\alpha}} \rho]_{{\bf Q}'} \ , 
\label{eq:deltarho3}
\end{multline}
having selected the cell ${\bf R}_{l}=0$ (without loss of generality for the argument given above regarding the independence of
Eq.~(\ref{eq:deltarho2}) on the cell index $l$) and having defined the 2D Fourier components: 
\begin{equation}
[\widetilde{\nabla}_{\boldsymbol{\tau}_{\alpha}} \rho]_{\bf Q}  \ = \ 
\int {\rm d}{\bf R} \ e^{i{\bf Q} \cdot {\bf R}} \ \widetilde{\nabla}_{\boldsymbol{\tau}_{\alpha}} \rho ({\bf R},d) \ .
\label{eq:deltarho4}
\end{equation}
Finally, assuming that the Green's function $G^{(0)}_{\bf Q}(z,d)$ is constant over the `vertical' support of the wavefunctions
(a good approximation especially in the important region of small $Q$), Eq.~(\ref{eq:matrixel4}) becomes:
\begin{multline}
\hspace*{-0.40cm}
\langle {\bf K'},n^{\prime}  | \delta E^{(\eta)}_{\rm H} | {\bf K},n \rangle  \approx  \\
\hspace*{-0.80cm}
   \sum_{{\bf g}' {\bf g} {\bf Q}} \ \delta_{{\bf K}'-{\bf K}+{\bf G}'-{\bf G},{\bf Q}}
                           \ \delta_{g_{z}',g_{z}}\ u^{(n')\ast}_{{\bf K}', {\bf g}'} \ u^{(n)}_{{\bf K}, {\bf g}} \\  
      \times \sum_{\alpha} \ \ e^{- i {\bf Q} \cdot \boldsymbol{\tau}_{\alpha}} \
        A^{(\eta)}_{\bf Q} {\bf e}^{(\eta, \alpha)}_{\bf Q} \cdot \widetilde{\nabla}_{\boldsymbol{\tau}_{\alpha}} \rho({\bf Q}) \ 
           G^{\rm (vac)}_{{\bf Q}}(d,d) \ .
\label{eq:matrixel5}
\end{multline}
Here, the Green's function $G^{(0)}_{\bf Q}(z,z')=e^{-Q|z-z'|}/(2 \epsilon_{0}Q)$ has been replaced by the Green's function, 
$G^{\rm (vac)}_{{\bf Q}}(z,z')$, of a system consisting of a 2D layer of dielectric constant $\epsilon^{(\infty)}_{\rm 2D}$ 
and thickness $h$. This may better mimic the geometry implicitly used in DFT calculations. 
We have also assumed that $G^{\rm (vac)}$ is almost constant over the thickness of the 2D layer
(once more, a good approximation for sufficiently small $Q$, the region in which screening is most effective):
\begin{equation}
G^{\rm (vac)}_{{\bf Q}}(d,d) \ = \ \frac{1}{2\epsilon_{\rm 2D}^{(\infty)} Q} \ \mathcal {G}^{\rm (vac)}_{\bf Q}(d,d) \ ,
\label{eq:Gvac}
\end{equation}
with:
\begin{equation}
\mathcal {G}^{\rm (vac)}_{\bf Q}(d,d) \ = \ 1 \ + \ 2 \lambda e^{-Qh} 
                         \frac{\lambda e^{-Qh} -1}{1- \lambda^{2}e^{-2Qh}} \ ,
\label{eq:Gvac2}
\end{equation}
and 
\begin{equation}
\lambda \ = \ \frac{\epsilon_{\rm vac}-\epsilon^{(\infty)}_{\rm 2D}}
                   {\epsilon_{\rm vac}+\epsilon^{(\infty)}_{\rm 2D}} \ .
\label{eq:lambda}
\end{equation}
We have made use of the assumed homogeneity and isotropy of the system on the $(x,y)$ plane, to express these Green's function as dependent
only on the magnitude $Q$ of ${\bf Q}$.\\

To summarize, assuming:
\begin{enumerate}
\vspace*{-0.10cm}
\item that the Hartree component dominates;
\vspace*{-0.15cm}
\item that for in-plane phonons the ionic displacement lies purely on the plane of the 2D crystal;
\vspace*{-0.15cm}
\item that the change of the charge caused by this displacement is localized at $z'=d=h/2$;
\vspace*{-0.15cm}
\item and that the Poisson Green's function is constant over the effective thickness $h$ of the 2D layer, 
\end{enumerate}
\noindent then, the electron-phonon matrix element can be written as:
\begin{multline}
\hspace*{-0.3cm}
\langle {\bf K} + {\bf Q},n^{\prime}  | \delta E^{(\eta)}_{\rm H} | {\bf K},n \rangle  \approx  
   \sum_{{\bf g} {\bf g}'}  \ u^{(n')\ast}_{{\bf K}+{\bf Q}, {\bf g}'} \ u^{(n)}_{{\bf K}, {\bf g}} 
      \sum_{\alpha} \ e^{- i {\bf Q} \cdot \boldsymbol{\tau}_{\alpha}} \\ \times
            A^{(\eta, \alpha)}_{\bf Q} {\bf e}^{(\eta, \alpha)}_{\bf Q} \cdot \widetilde{\nabla}_{\boldsymbol{\tau}_{\alpha}} \rho({\bf Q})  
                  \ G^{\rm (vac)}_{\bf Q}(d,d) \ ,
                  \label{eq:matrixel6}
\end{multline}
having ignored {\it Umklapp} contributions (that is, having replaced ${\bf Q}+{\bf G}-{\bf G}'$ by ${\bf Q}$), 
since the Poisson Green's function $G^{\rm (vac)}_{\bf Q}(d,d)$ decays quickly with $Q$.\\ 

The important result of this long discussion is that, for a given wave vector ${\bf Q}$, the matrix element depends linearly on the Poisson Green's function
$G^{\rm (vac)}_{\bf Q}(d,d)$. Therefore, to estimate the matrix elements in the dielectric environment described above in Sec.~\ref{sec:system}, 
we may simply replace the `vacuum' Poisson Green's function $G^{\rm (vac)}_{\bf Q}(d,d)$ with the Green's function for the Poisson equation with the
boundary conditions dictated by the double-gate geometry of Sec.~\ref{sec:system}. This can be calculated using trivial but cumbersome algebra and, 
for the general double-gate geometry, 
$G^{\rm (env)}_{\mathbf{Q},\omega}(d,d)$ is given by:
\begin{equation}
    G^{\rm (env)}_{\mathbf{Q},\omega}(d,d) = \frac{1}{2\epsilon_{2D}(Q,\omega)Q}\mathcal{G}_{Q,\omega}^{\rm (env)}(d,d) \ ,
\label{eq:Genv} 
\end{equation}
where:
\begin{multline}
    \mathcal{G}_{Q,\omega}^{\rm (env)}(d,d) = \\
       1 + e^{-Qh}\frac{2\lambda_{b}(Q,\omega)\lambda_{t}(Q,\omega)e^{-Qh}-\lambda_{b}(Q,\omega)-\lambda_{t}(Q,\omega)}
                     {1-\lambda_{b}(Q,\omega)\lambda_{t}(Q,\omega)e^{-2\mathbf{Q}h}} \ .
\label{eq:Genv1}
\end{multline}
\noindent The quantities $\lambda_{\rm b}(Q,\omega)$ and $\lambda_{\rm t}(Q,\omega)$ are given by:
\begin{equation}
    \lambda_{\rm b}(Q,\omega) = 
       \frac{\epsilon_{\rm box}(\omega)\coth(Q,t_{b})-\epsilon_{\rm 2D}(Q,\omega)}{\epsilon_{\rm box}(\omega)\coth(Q,t_{b})+\epsilon_{\rm 2D}(Q,\omega)} \ ,
\label{eq:lambda1}
\end{equation}
\begin{equation}
    \lambda_{\rm t}(Q,\omega) = 
       \frac{\epsilon_{\rm tox}(\omega)\coth(Q,t_{t})-\epsilon_{\rm 2D}(Q,\omega)}{\epsilon_{\rm tox}(\omega)\coth(Q,t_{t})+\epsilon_{\rm 2D}(Q,\omega)} \ ,
\label{eq:lambda2}
\end{equation}
\noindent with $\epsilon_{\rm box}(\omega)$ and $\epsilon_{\rm tox}(\omega)$ given by Eqs.~(\ref{eq:epsbox}) and (\ref{eq:epstox}) and $\epsilon_{2D}(Q,\omega)$ given by Eq.~(\ref{eq:eps2D}) (or Eq.~(\ref{eq:eps2D1}) in the simpler small-$Q$ approximation). 
Therefore, for a given scattering process involving a phonon of branch $\eta$ and wave vector $\mathbf{Q}$, we can account for the presence of the 
different dielectric environment and free-carrier screening by simply rescaling the scattering rates calculated using DFT/EPW by the squared ratio:
\begin{equation}
    \left|\frac{\mathcal{G}_{Q,\omega^{(\eta)}_{Q}}^{(env)}(d,d)}{\mathcal{G}_{Q}^{\rm (vac)}(d,d)}\right|^{2} \ .
\label{eq:scale}
\end{equation}

This can be easily implemented in Monte Carlo simulations: We use the `bare' scattering rates computed for free-standing layers but, after a collision 
involving a TMD phonon with wave vector ${\bf Q}$ and frequency $\omega^{(\eta)}_{Q}$, we accept or reject the collision with a probability distribution
given by Eq.~(\ref{eq:scale}).\\

\subsection{Electron-IPP scattering}
\label{sec:rph}

\subsubsection{Secular equation, IPP dispersion, and scattering potential for the double-gate geometry}
\label{sec:IPP}

In order to obtain the dispersion of the hybrid IPPs, we must first consider the potential associated with these excitations. Here we follow the procedure 
described in Ref.~\cite{Fischetti_2001} for the gate-insulator/Si-inversion-layer 
system and in Refs.~\cite{Ong_2013} and~\cite{Ong_2012} for supported graphene and supported and gated graphene, respectively.
For the double-gate geometry described in Sec.~\ref{sec:system}, the potential has the form:
\vspace*{-0.15cm}
\begin{equation}
    \phi_{Q,\omega}(z) = 
    \begin{cases}
    a e^{Qz}+Ae^{-Qz}\;\;\;\;\;(-t_{\rm b}\leq z\leq0)\\
    be^{Qz}+de^{-Qz} \;\;\;\;\;(0<z\leq h)\\
    de^{Qz}+fe^{-Qz} \;\;\;\;\;(h<z\leq t_{\rm t}+h)\\
 \end{cases}
\label{eq:pot_DG}   
 \end{equation}
Imposing the continuity of the components of the electric field, $\mathbf{E}_{\parallel}$, and of the displacement field, $D_{z}$, parallel and normal, respectively,  to the plane of the interfaces, and setting $\phi_{Q,\omega}(- t_{\rm b}) = \phi_{Q,\omega}(t_{\rm t} + h)= 0$, as required by the presence
of the metal gates, we obtain a linear system of 6 equations in 6 unknowns ($A$, $a$, $b$, $c$, $d$, and $f$)  
that admits a nontrivial solution only if the determinant of the coefficients vanishes. This leads to the secular equation
(omitting for simplicity the dependence of the dielectric functions on $Q$ and $\omega$):
\begin{multline}
\label{eqn:seculardg}
        \epsilon_{\rm 2D}[(\hat{\epsilon}_{\rm 2D}-\hat{\epsilon}_{\rm box})e^{-Qh}-(\hat{\epsilon}_{\rm 2D}+\hat{\epsilon}_{\rm box})e^{Qh}\\
        -\epsilon_{\rm tox}[(\hat{\epsilon}_{\rm 2D}
        -\hat{\epsilon}_{\rm box})e^{-Qh}-(\hat{\epsilon}_{\rm 2D}+\hat{\epsilon}_{\rm box})e^{Qh}\; {\rm coth}(Qt_{\rm t}) = 0 \ ,
\end{multline}
\noindent where $\hat{\epsilon}_{\rm 2D}$ = $\epsilon_{\rm 2D} \sinh(Qt_{\rm b})$ and $\hat{\epsilon}_{\rm box}$ = $\epsilon_{\rm box} \cosh(Qt_{\rm b})$.
This is an algebraic equation of 8$^{\rm th}$ degree in $\omega^{2}$
whose 8 solutions for a fixed $Q$ give the dispersion $\omega_{Q}^{(i)}$ (with $i=1,8$). Note that, 
with some straightforward but laborious algebra, this equation can be rewritten in a form that exhibits explicitly its symmetry under exchange 
of the top and bottom dielectrics:
\vspace*{-0.25cm}
\begin{multline}
\label{eqn:secular_symmetric}
   \epsilon_{\rm 2D}^{2} \tanh(Qt_{\rm b}) \tanh(Qt_{\rm t}) +
      \epsilon_{\rm box} \epsilon_{\rm tox} \tanh(Qh) \\
        + \epsilon_{\rm 2D} \epsilon_{\rm tox} \tanh(Qt_{\rm b}) + 
           \epsilon_{\rm 2D} \epsilon_{\rm box} \tanh(Qt_{\rm t}) = 0 \ .
\end{multline}

Once the secular equation is solved, for each mode $i$, the potential that solves the homogeneous system can be expressed in terms of an arbitrary multiplication constant. Solving for $a$, $b$, $c$, $d$, and $f$ in terms of $A$, we have:
\begin{widetext}
\begin{subequations}
\begin{align}
a = A \ e^{-2Qt_{b}}            \label{eq:Asystema} \\
b = A \ \frac{1}{2}\left[1-e^{-2Qt_{b}}+\frac{\epsilon_{\rm box}(\omega)}{\epsilon_{\rm 2D}(Q,\omega)}(1+e^{-2Qt_{b}})\right] \equiv A \ \mathcal{B}_{Q,\omega},
                             \label{eq:Asystemb} \\  
c = A \ \frac{1}{2}\left[1-e^{-2Qt_{b}}-\frac{\epsilon_{\rm box}(\omega)}{\epsilon_{\rm 2D}(Q,\omega)}(1+e^{-2Qt_{b}})\right] \equiv A \ \mathcal{C}_{Q,\omega},
                             \label{eq:Asystemc} \\  
d = A \ \frac{\epsilon_{\rm 2D}(Q,\omega) (1 - e^{-2Qt_{\rm b}} ) \sinh(Qh)
         + \epsilon_{\rm box}(\omega) ( 1 + e^{-2Qt_{\rm b}} ) \cosh(Qh)}{\epsilon_{\rm tox}(\omega)[e^{Qh}+e^{-Q(h-2t_{\rm t})}]} \equiv 
                                                           A \ \mathcal{D}_{Q,\omega} \ ,
                            \label{eq:Asystemd} \\
f = A \ \frac{ \epsilon_{\rm 2D}(Q,\omega) ( 1 - e^{-2Qt_{\rm b}}) \sinh(Qh)
         + \epsilon_{\rm box}(\omega) ( 1 + e^{-2Qt_{\rm b}}) \cosh(Qh)}{\epsilon_{\rm tox}(\omega)[e^{-Qh}+e^{Q(h-2t_{\rm t})}]} \equiv  
                                                  A \ \mathcal{F}_{Q,\omega} \ ,
                            \label{eq:Asysteme} 
\end{align}     
\end{subequations}
\end{widetext}
having made use the secular equation to write the last two expressions. These equations define implicitly the functions 
$\mathcal{B}_{Q,\omega}, ..., \mathcal{F}_{Q,\omega}$.

In order to determine the constant $A$ (all-important, since it determines the overall strength of the interaction), one could quantize these excitations
following the usual canonical quantization procedure and the magnitude of $A$ would be determined by the canonical commutation rules. However, here we follow 
the simpler but equivalent procedure of Refs.~\cite{Fischetti_2001,Ong_2012,Ong_2013}, based 
on Stern and Ferrell~\cite{EAStern_1960}: For each hybrid mode $i$, we calculate the total energy of the field 
(time-averaged and including self-energy terms) in terms of $A$ and set it equal to the ground-state energy $\hbar \omega^{(i)}/2$. 
However, in order to consider separately the contribution $\Phi^{(\alpha)}(\omega)$ due to each bare phonon $\alpha$ (= TO1, TO2, TO3, TO4, or ZO, 
the `phonon content' discussed below and in Refs.~\cite{Ridley_book,Varga_1965,Kim_1978,Fischetti_2001,Ong_2013}), we consider separately the time-averaged
energy $\langle W^{(\alpha)}_{Q,\omega_{Q}^{(i)}} \rangle$ due to each bare mode $\alpha$ and set it equal to a fraction $\Phi^{(\alpha)}(\omega_{Q}^{(i)})$ 
of the oscillator ground-state energy:
\begin{multline}
\frac{2}{\Omega} \ \left \langle W^{(\alpha)}_{Q,\omega_{Q}^{(i)}} \right \rangle \
           = \ \frac{1}{\Omega} \int_{-\infty}^{t_{\rm t}} {\rm d}z \ \rho_{Q,\omega_{Q}^{(i)}}(z) \ \phi^{(\alpha)}_{Q,\omega_{Q}^{(i)}}(z) \\ 
                    = \ \frac{1}{2} \ \hbar \omega_{Q}^{(i)} \Phi^{(\alpha)}(\omega_{Q}^{(i)})  \ ,
\label{eq:W1}
\end{multline}
where $\Omega$ is the normalization area. 

Proceeding as in Refs.~\cite{Fischetti_2001,Ong_2012,Ong_2013} 
and taking as scattering potential the potential at $z=0$, we obtain 
for the scattering strength relative to the content of the `bare' phonon $\alpha$ of the hybrid mode $i$:
\begin{widetext}
\begin{equation}
\left \vert {\mathcal{A}}^{(\alpha)}_{Q,\omega_{Q}^{(i)}} \right \vert^{2}  \ = \ \frac{\hbar \omega_{Q}^{(i)}}{2Q} \ \left ( 1 - e^{-2Qt_{\rm b}}  \right )^{2} 
         \Phi^{(\alpha)}(\omega^{(i)}_{Q})  
             \left \vert \frac{1}{\epsilon_{\rm TOT}^{(\alpha,\rm high)}(Q,\omega_{Q}^{(i)})} 
                                    - \frac{1}{\epsilon_{\rm TOT}^{(\alpha,\rm low)}(Q,\omega_{Q}^{(i)})} \right \vert \ .                                                      
\label{eq:amplitude_DG}
\end{equation}
where $\epsilon^{(\alpha,\rm low)}_{\rm TOT}(Q,\omega)$ and $\epsilon^{(\alpha,\rm high)}_{\rm TOT}(Q,\omega)$ are the `total' dielectric functions
of the system assuming that mode $\alpha$ responds fully (`low') or does not respond (`high').
They are given by:
\begin{multline}
\epsilon_{\rm TOT}(Q,\omega) = 
   \left [ \widetilde{\epsilon_{\rm box}}(\omega) (1-e^{-2Qt_{\rm b}}) - \widetilde{\epsilon_{\rm 2D}}(Q,\omega) 
          ( \mathcal{B}_{Q,\omega} - \mathcal{C}_{Q,\omega} ) \right ] (1-e^{-2Qt_{\rm b}})  \ +  \\
        \left [ \widetilde{\epsilon_{\rm 2D}}(Q,\omega)  ( \mathcal{B}_{Q,\omega} e^{Qh} - \mathcal{C}_{Q,\omega} e^{-Qh} )
      - \widetilde{\epsilon_{\rm tox}}(\omega) ( \mathcal{D}_{Q,\omega} e^{Qh} - \mathcal{F}_{Q,\omega} e^{-Qh} ) \right ]
          ( \mathcal{B}_{Q,\omega} e^{Qh} + \mathcal{C}_{Q,\omega} e^{-Qh} ) \ .  
\label{eq:epstot}            
\end{multline} 
\end{widetext}
Here the functions $\widetilde{\epsilon_{\rm box}}(\omega)$, $\widetilde{\epsilon_{\rm tox}}(\omega)$, and $\widetilde{\epsilon_{\rm 2D}}(Q,\omega)$
are calculated considering the full response of phonon $\alpha$ (`low') or assuming that mode $\alpha$ does not respond (`high'). 
This is done by setting $\omega=0$ or $\omega \rightarrow \infty$, respectively, in the expression representing the contribution of phonon $\alpha$ to the 
polarizability. For example, if $\alpha$ = TO2 (the high-frequency phonon of the bottom oxide), we set:
\vspace*{-0.10cm}
\begin{equation}
 \epsilon_{\rm box}^{\rm (TO2,high/low)}(\omega) =  \epsilon_{\rm box}^{\rm (\infty/mid)} + 
                      [ \epsilon_{\rm box}^{(0)} - \epsilon_{\rm box}^{\rm (mid)} ] \
                         \frac{\omega_{\rm TO,1}^{2}}{\omega_{\rm TO,1}^{2} - \omega^{2}} \ .
\label{eq:epshi}
\end{equation} 

As mentioned above, the quantity $\Phi^{(\alpha)}(\omega^{(i)}_{Q})$ appearing in 
Eqs.~(\ref{eq:W1}) and (\ref{eq:amplitude_DG}) denotes the content of the (uncoupled) phonon 
mode $\alpha$ of the hybrid mode $i$. To obtain this quantity we follow Refs.~\cite{Ridley_book,Varga_1965,Kim_1978}, as described also in Ref.~\cite{Fischetti_2001} and \cite{Ong_2013}. 

Equation~(\ref{eq:amplitude_DG}) reflects the fact that we have chosen the potential at $z=0$ as scattering
potential. The full symmetry of the system when $t_{\rm b} = t_{\rm t}$ may be restored by using, instead, the potential at $z=d=h/2$.
However, this would not result in any significant change in the long-wavelength limit, $Q < 1/d$, the range of wavelengths in which the 
IPP potential is strongest. This is consistent with the approximations also made in Eqs.~(\ref{eq:rate3}) below and 
we shall show in Sec.~\ref{sec:IPP-scattering} that our results confirm the validity of the long-wavelength approximation. 

\subsubsection{Landau damping}
\label{sec:Landau}

The entire discussion has been based so far on the assumption that all the decoupled excitations of interest, electrons/holes, plasmons, and phonons,
are pure eigenstates of the total Hamiltonian of the system. Clearly, this ignores the fact that optical phonons decay into acoustic phonons 
via two-phonon~\cite{Klemens_1966} or many-phonon processes~\cite{ziman2001electrons,Egorov_1995,Berke_1988,feng2016quantum} with a rather short
lifetime of the order of ps; that acoustic phonons also decay, although with a much longer lifetime (often more than hundreds 
of ns~\cite{ziman2001electrons,Berke_1988}, shorter -- but still hundreds of ps -- in MoS$_2$ monolayers~\cite{Guo_2018}), 
in addition to decaying via electron-phonon interactions and scattering with boundaries/interfaces. Even more important is the fact that this assumption
ignores also the decay of plasmons into single-particle excitations via Landau damping~\cite{landau196561,pines1961collective,pines1962approach,bohm1953collective,tsu1967landau}. 
As discussed above (Sec.~\ref{sec:system}),
for wave vectors ${\mathbf Q}$ such that 
$E(K_{\rm F} - Q) - E(K_{\rm F}) \le \hbar \omega_{\rm P}(Q) \le E(K_{\rm F} + Q) - E(K_{\rm F})$ 
(assuming isotropic dispersions for electrons and plasmons), plasmons cease to exist as well-defined eigenstates of the Hamiltonian and 
decay into single particle excitations.

The approach followed by Hauber and Fahy~\cite{Hauber_2017} accounts for Landau damping correctly in the case of the plasmon/optical-phonon coupling in 
bulk polar semiconductors and of the bottom-insulator/MoS$_{2}$/top-insulator system (namely the vacuum/MoS$_{2}$/vacuum, hBN/MoS$_{2}$/hBN, 
HfO$_{2}$/MoS$_{2}$/SiO$_{2}$, and vacuum/MoS$_{2}$/SiO$_{2}$ stacks). They have expressed the scattering rate in terms of the imaginary part of the total
dielectric function, as it follows from the fluctuation-dissipation theorem~\cite{Quinn_1958,Quinn_1962,Penn_1987,Sandborn_1998}. Such a procedure can
become quite cumbersome in complicated geometries. Therefore, we have reverted to our 'brute-force' approximation used in the past~\cite{Fischetti_2001}, 
but improved following Hauber and Fahy's suggestion: Rather than considering unscreened phonons in the Landau-damped (single-particle) region,   
$Q \ge Q_{\rm LD}$ (where the Landau-damping cut-off wave vector, $Q_{\rm LD}$, has been already defined implicitly in Eq.~(\ref{eq:LD2}) as being such that 
$E(K_{\rm F} + Q_{\rm LD}) - E(K_{\rm F}) = \hbar \omega_{\rm P}(Q_{\rm LD})$), 
we have replaced the dynamic long-wavelength term $1-\omega_{\rm P}(Q)^{2}/\omega^{2}$ in Eq.~(\ref{eq:eps2D1}) with the static 
Thomas-Fermi expression $1+Q_{\rm TF}^{2}/Q^{2}$. We denote by $Q_{\rm TF}$ the screening wave vector given by: 
\begin{equation}
Q_{\rm TF} = \frac{e^{2} n}{2 \epsilon_{\rm 2D, v} k_{\rm B}T } \ ,
\label{eq:QTFnd}
\end{equation}
in the non-degenerate limit ($n$ is carrier density), or
\begin{equation}
Q_{\rm TF} = \frac{e^{2} m^{\ast}g}{2 \pi \hbar^{2} \epsilon_{\rm 2D, v} } \ ,
\label{eq:QTFdeg}
\end{equation}
in the degenerate limit ($g$ is the valley degeneracy).  
This accounts approximately for the static response of the free carriers when the frequency of the perturbation is smaller 
than the plasma frequency; that is, when $\omega_{\rm P}(Q) \gg \omega^{(i)}_{Q}$. 
In this region of wavelengths, the resulting secular equation yields only phonon-like branches, 
6 of them in the DG geometry we have considered, each of them screened statically by the 2DEG. However, as we have already stressed before, electrons couple only
with the components of four of the `bare' modes, since the contribution to the due to the two modes arising from the TMD ZO phonons at the top and bottom
interfaces is anti-symmetric~\cite{Zhang_2020}, so that its matrix elements vanish in mirror-symmetric ($\sigma_{\rm h}$-symmetric) monolayers~\cite{Fischetti_2016}. 

\subsubsection{Electron-IPP scattering rates}
\label{sec:IPPrate}

From Eqs.~(\ref{eq:pot_DG}), (\ref{eq:Asystema})-(\ref{eq:Asysteme}), and (\ref{eq:amplitude_DG}), the IPP scattering potential needed to calculate
the scattering rates has the form:
\begin{multline}
\phi_{Q,\omega}({\bf r}) \ = e^{i {\bf Q} \cdot {\bf R}} {\mathcal{A}}_{Q,\omega} \times \\ \left \{
  \begin{array}{ll}
   e^{-2Qt_{\rm b}}        \ e^{Qz} \ + \                          \ e^{-Qz}  & (-t_{\rm b} < z \le 0) \\  
  {\mathcal{B}}_{Q,\omega} \ e^{Qz} \ + \ {\mathcal{C}}_{Q,\omega} \ e^{-Qz} & (0 < z \le h) \\
  {\mathcal{D}}_{Q,\omega} \ e^{Qz} \ + \ {\mathcal{F}}_{Q,\omega} \ e^{-Qz} & (h < z \le t_{\rm t}) \\
  \end{array}
  \right. \ .
\label{eq:potential2}  
\end{multline}

Using Fermi's golden rule, the scattering rate for an electron or hole in band $n$ and 2D wave vector ${\bf K}$ to emit or absorb a hybrid excitation 
of mode $i$ with wave vector ${\bf Q}$ and frequency $\omega^{(i)}_{Q}$ can be written as:
\begin{multline}
\frac{1}{\tau^{(i)}_{n}({\bf K})} \ \approx \  \\ \frac{2 \pi}{\hbar} \ \sum_{n' {\bf K}' {\bf Q}} \
         \left \rvert \int_{\Omega} {\rm d} {\bf r} \ \psi^{\ast}_{n' {\bf K}'}({\bf r}) \
                                                       e \ \phi_{ Q, \omega^{(i)}_{Q} } ({\bf r}) \ \psi_{n {\bf K}} ({\bf r}) \right \rvert^{2} \\ \times
          \ \left \{ \begin{array}{cc} N[\omega^{(i)}_{\bf Q}] \\ 1+N[\omega^{(i)}_{\bf Q}] \end{array} \right \} \
          \ \delta ( E_{n{\bf K}} - E_{n'{\bf K}'} \pm \hbar \omega^{(i)}_{Q} ) \ ,
\label{eq:rate1}
\end{multline}
where $\Omega$ is the normalization area and $N(\omega)$ is the equilibrium Bose-Einstein distribution function. The upper/lower term refers to absorption/emission of each hybrid mode. 
Using the Bloch form of the wavefunctions, $\psi_{n {\bf K}} ({\bf r})=e^{i {\bf K} \cdot {\bf R}} \ u^{(n)}_{\bf K}({\bf r})/\Omega^{1/2}$,
we can rewrite the matrix element appearing in this expression as:
\begin{equation}
\frac{1}{\Omega} \ \int_{\Omega} {\rm d}{\bf r} \ u^{(n')\ast}_{{\bf K}'} ({\bf r}) \ e^{i {\bf K}' \cdot {\bf R}} \ 
                                                  e \phi_{Q, \omega^{(i)}_{Q} }(z) \ e^{i {\bf Q} \cdot {\bf R} } \
                                                  u^{(n)}_{{\bf K}}({\bf r}) \ e^{i {\bf K} \cdot {\bf R}} \ .
\label{eq:rate2}
\end{equation}

As we have already mentioned, in principle the Bloch functions and their energy $E_{n,{\bf k}}$ depend on the 3D wave vector ${\bf k}=({\bf K},k_{z})$. 
However, we ignore the dependence of $E_{n,{\bf k}}$ and $\psi_{n,{\bf K}}({\bf r})$ on the out-of-plane component of the wave vector, $k_{z}$ and also 
assume (as it would be correct only in the limit $h \rightarrow 0$) that  
the wavefunctions are localized on the plane of the 2D layer, so that
$u^{(n)}_{\bf K}({\bf r}) \approx U^{(n)}_{\bf K}({\bf R}) \delta(z)$, and Eq.~(\ref{eq:rate2}) becomes:
\begin{equation}
\frac{1}{\Omega} \ e {\mathcal{A}}_{Q,\omega^{(i)}_{Q}} 
        \ \int_{\Omega} {\rm d}{\bf R} \ U^{(n')\ast}_{{\bf K}'}({\bf R}) \ U^{(n)}_{{\bf K}}({\bf R}) 
                               e^{i ({\bf K}-{\bf K}'+{\bf Q}) \cdot {\bf R}} \ .
\label{eq:rate3}
\end{equation}

With the exception of the matrix element for the interaction with the ZO phonon-component of the potential, following the results of 
Ref.~\cite{Zhang_2020}, treating the TMD wavefunctions as delta-functions, $\delta(z-d)$, 
should be a reasonable approximation also when assuming a non-zero thickness, but assuming a constant potential 
(along the out-plane direction $z$) in the 2D layer and wavefunctions similarly constant in the layers and zero outside the layer.
At small $Q$, only intravalley scattering occurs and, in this case, the only error originates from ignoring
the overlap integral between the initial and final states. For intravalley processes, this is a minor effect. On the
contrary, it may be an important effect for intervalley processes that may occur at large $Q$. However, in this case the scattering rate 
is small, because of the $1/Q$ dependence of the squared matrix elements.

Following the `usual' procedure (see Eqs.~(12.3)-(12.9) of Ref.~\cite{Mybook}), setting ${\bf R} = {\bf R}_{l} + {\bm x}$ where ${\bf R}_{l}$ 
are the lattice vectors and ${\bm x}$ spans a unit cell, splitting the integral over the entire volume into a sum over cells times 
an integral over a single cell with area $\Omega_{\rm cell}$, this becomes: 
\begin{multline}
\sum_{\bf G} \ e {\mathcal{A}}_{|{\bf Q}+{\bf G}|,\omega^{(i)}_{Q}} \ \delta^{(2)}({\bf K}-{\bf K}' + {\bf Q}) \\
         \times \frac{1}{\Omega_{\rm cell}} \ \int {\rm d}{\bm x} \ U^{(n')\ast}_{{\bf K}'}({\bm x}) \ U^{(n)}_{{\bf K}}({\bm x}) 
                               \ e^{i {\bf G} \cdot {\bm x}} \ .
\label{eq:rate4}
\end{multline}

Finally, assuming that the overlap integral between the initial and final Bloch functions is unity (that is, ignoring `overlap-factor' effects,
usually negligible when ${\bf K}$ and ${\bf K}'$ differ only by a small wave vector ${\bf Q}$), 
ignoring {\it Umklapp} processes, thanks to the dependence of the matrix element on $1/Q$, in the infinite-volume normalization we finally have:
\vspace*{-0.750cm}
\begin{widetext}
\begin{multline}
\frac{1}{\tau^{(i)}_{n}({\bf K})} \ \approx \ \frac{2 \pi}{\hbar} \ \sum_{n' \alpha} \ \int \frac{{\rm d} {\bf Q}}{(2 \pi)^{2}} \
        \frac{e^{2} \hbar \omega^{(i)}(Q)}{2Q} \ \Phi^{(\alpha)}[\omega^{(i)}(Q)] \
                          \left \vert \frac{1}{\epsilon_{\rm TOT}^{(\alpha,\rm high)}[Q,\omega^{(i)}(Q)]} 
                                    - \frac{1}{\epsilon_{\rm TOT}^{(\alpha,\rm low)}[(Q,\omega^{(i)}(Q)]} \right \vert \\
                 \times \ \left \{ \begin{array}{cc} N[\omega^{(i)}(Q)] \\ 1+N[\omega^{(i)}(Q)] \end{array} \right \} \
                        \ \delta [ E_{n}({\bf K}) - E_{n'}({\bf K}+{\bf Q}) \pm \hbar \omega^{(i)}(Q) ] \ .
\label{eq:rate5}
\end{multline}
\end{widetext}

It can be shown in general that the function $\epsilon_{\rm TOT}(Q,\omega)$ defined by Eq.~(\ref{eq:epstot}) 
(with the dielectric functions $\widetilde{\epsilon}$ replaced by their normal expressions $\epsilon$) is identical
to the left-hand side of the secular equation, Eq.~(\ref{eqn:seculardg}). Moreover,    
a little algebra shows that, in the limit $h \rightarrow 0$, Eq.~(\ref{eq:epstot}) implies 
$\epsilon_{\rm TOT}(Q,\omega) \rightarrow (1-e^{-2Qt_{\rm b}})^{2}[\epsilon_{\rm box}(\omega) \coth(Qt_{\rm b})+\epsilon_{\rm tox}(\omega) \coth(Qt_{\rm t}]$
which, in turn, tends to $\epsilon_{\rm box}(\omega) +\epsilon_{\rm tox}(\omega)$ in the limit of infinitely thick insulators (that is, 
$t_{\rm b}, t_{\rm t} \rightarrow \infty$). These are the `usual' expressions obtained in Refs.~\cite{Wang_1972} and \cite{Ong_2013}.    
Therefore, in these latter limits, for a single interface between two dielectrics, we recover the usual expression 
for the `Fr\"{o}hlich' term inside the absolute value in Eq.~(\ref{eq:rate5}):
\vspace*{-0.25cm}
\begin{multline}
 \left \vert \frac{1}{\epsilon_{\rm box}^{(\alpha,\rm high)}(Q,\omega_{Q}^{(i)})+\epsilon_{\rm tox}^{(\alpha,\rm high)}(Q,\omega_{Q}^{(i)})} \right. \\
   \left.  - \frac{1}{\epsilon_{\rm box}^{(\alpha,\rm low )}(Q,\omega_{Q}^{(i)})+\epsilon_{\rm tox}^{(\alpha,\rm low )}(Q,\omega_{Q}^{(i)})} \right \vert \ .
\label{eq:rate5a}
\end{multline}
Explicit analytic calculations for a simple Si/SiO$_2$ system show that this expression yields the same scattering 
rates obtained by Hauber and Fahy~\cite{Hauber_2017} when accounting for a slightly different definition of $\epsilon_{\rm TOT}(Q,\omega)$.
  
\subsection{Full-band Monte Carlo simulations}
\label{sec:MC}
The low-field carrier mobility has been calculated using the Monte Carlo method described in-detail in our previous work presented in   Refs.~\cite{gaddemane2021limitations,gaddemane2018theoretical}. To provide here just the basic features of the numerical method, the band structure, scattering
rates and phonon spectra discussed above are interpolated and tabulated on a fine {\bf k}-mesh, described in Sec.~\ref{sec:abinitio}. that spans the entire Brillouin zone. As discussed in Refs.~\cite{Sohier_2018}, \cite{Ponce_2018}, and \cite{gaddemane2021limitations}, the extremely fine mesh we use,  
mentioned in Sec.~\ref{sec:abinitio} (namely: a 201×201×1 mesh covering a rectangular section that inscribes the triangular irreducible wedge of the hexagonal first Brillouin zone) is required to account correctly and accurately not only for energy conservation but also for 
the anisotropy and the strong nonparabolicity of the band structure around the conduction-band
minimum (or minima) and valence-band maximum. (Additional details about the algorithms used in our numerical implementation of the Monte Carlo method
are given in the `Supplemental material' of Ref.~\cite{gaddemane2018theoretical}.) 
A bilinear interpolation is used to obtain information at arbitrary points in the Brillouin zone. As already mentioned, scattering 
with the flexural acoustic and optical phonons (the out-of-plane ZAs and ZOs) is ignored, since the horizontal mirror ($\sigma_{\rm h}$) symmetry exhibited 
by the 2H TMDs we have considered forbids these interactions at first order. Second-order (two-phonon) processes, while important in 
graphene~\cite{Mariani_2008,VonOppen_2009, Bolotin_2008,Castro_2010,Gornyi_2012}, are negligible in the materials of interest here~\cite{Rudenko_2019}.
The carrier mobility, $\mu$, is extracted from the zero-field diffusion constant $D$ via the Einstein relation, generalized in the case of degenerate
TMDs, rather than from the velocity- field characteristics, since the former is less affected by stochastic noise~\cite{jacoboni1983monte}. The simulations
are performed using a synchronous ensemble of 500 particles for simulation times of about 10~ps, as required to reach steady state and a sufficiently
accurate statistics.
\vspace*{-0.50cm}
\section{Results and Discussion}
\label{sec:results}

In this section, we present our results -- obtained using the full-band Monte Carlo method described above -- regarding the effect of dielectric screening 
and IPP scattering on the transport properties of the double-gated monolayer TMD system discussed in Section \ref{sec:system}. We discuss the carrier mobility
in several supported and/or gated monolayer TMDs, all 
assumed to be supported by SiO$_2$. Examples of the band structure and phonon spectra for various TMDs considered here can be found in Ref.~\cite{gaddemane2021limitations}. Table 3 of this reference lists also the all-important values of the energy separation between the conduction-band minima 
at the symmetry point K and the minima of the satellite Q-valleys obtained using the ONCV pseudopotentials and the GGA-PBE exchange-correlation functional.

Tables \ref{tab:para} and \ref{tab:par} list the physical parameters used in our calculations. The dielectric constant and layer thickness of monolayer TMDs 
are taken from Ref.~\cite{laturia2018dielectric}. The transverse optical (TO) phonon frequencies and the dielectric constant for the polar insulators are 
obtained from Ref.~\cite{Fischetti_2001}. We have used the 'effective' dielectric constant for the polar insulators in our calculations, as suggested
by~\cite{Hauber_2017}. 
\begin{table}[tb] 
\caption{Physical parameters used to calculate IPP scattering in the 2H TMD monolayers.}
\vspace*{0.5cm}
\begin{tabular}{lllllll}
\hline
\hline
Parameter                                       & MoS$_{2}$ & MoS$_{2}$ &  MoTe$_{2}$ & WS$_{2}$ &  WSe$_{2}$ &  WTe$_{2}^{\ d}$ \\
\hline
$\hbar \omega_{\rm ZO}$ (meV)$^{a}$             &    34.2   &  20.14    &  28.17      &   35.6   &   20.51    &   14.6        \\
$t_{\rm 2D}$ (nm)$^{b}$                         &   0.612   &  0.649    &   0.71      &  0.612   &   0.648    &   0.422       \\
$\epsilon^{(0)}$ ($\epsilon_{0}$)$^{^{b}}$      &     9.8   &  11.19    &  14.02      &   9.34   &   10.74    &   10.42       \\
$\epsilon^{(\infty)}$ ($\epsilon_{0}$)$^{^{b}}$ &     9.69  &  10.99    &  13.24      &   9.24   &   10.64    &   10.32       \\
$m^{\ast}_{\rm e}$ ($m_0)$$^{^{c}}$             &     0.5   &   0.55    &   0.65      &   0.3    &   0.6      &   0.4         \\
$m^{\ast}_{\rm h}$ ($m_0)$$^{^{c}}$             &     0.6   &   0.62    &   0.95      &   0.4    &   0.35     &   0.5         \\  
\hline
\hline
\end{tabular}\\
{\small \flushleft $^{a}$ From our DFT calculations\\
$^{b}$ From Ref.~\cite{laturia2018dielectric}\\
\hspace*{-0.0cm}$^{c}$ Used only to compute $\omega_{\rm P}(Q)$ and screening wavelength\\
\hspace*{-6.4cm}$^{d}$ See note~\cite{bnote3}}
\label{tab:para}
\end{table}
\begin{table}[tb] 
\caption{Physical parameters of the polar insulators used to calculate IPP scattering.}
\vspace*{0.5cm}
\begin{tabular}{lllllll}
\hline
\hline
Parameter                                &  HfO$_{2}$  &  SiO$_{2}$  &  Al$_{2}$O$_{3}$ &    AlN    &   h-BN   &   ZrO$_{2}$  \\
\hline
$\hbar \omega_{\rm TO1}$ (meV)$^{a}$     &   12.4      &    55.6     &    48.18         &    81.4   &   92.5   &     16.67    \\
$\hbar \omega_{\rm TO2}$ (meV)$^{a}$     &   48.35     &   138.1     &    71.41         &    88.55  &  170.1   &     57.7     \\
$\epsilon^{(0)} (\epsilon_{0})^{a}$      &   22.0      &     3.9     &    12.53         &     9.14  &    5.1   &     24.0     \\  
$\epsilon^{\rm (mid)}(\epsilon_{0})^{a}$ &    6.58     &     3.05    &     7.27         &     7.35  &    4.45  &      7.75    \\
$\epsilon^{(\infty)}(\epsilon_{0})^{a}$  &    5.03     &     2.50    &      3.2         &     4.8   &    3.88  &      4.0     \\
$t_{\rm ox}$ (nm)$^{b}$                  &    4.2      &     0.7     &      2.25        &     1.64  &    0.916 &      4.31    \\
\hline
\hline
\end{tabular}\\
{\small \flushleft $^{a}$ From Ref.~\cite{Fischetti_2001}\\
\hspace*{-0.048cm} $^{b}$ Corresponding to an equivalent SiO$_2$ thickness of 0.7~nm.\\ 
\hspace*{0.300cm}The top-insulator thickness was varied for some of the\\ 
\hspace*{-3.275cm}calculations shown in Sec.~\ref{sec:IPP-scattering}.}
\label{tab:par}
\end{table}
\begin{figure}[h!]
\includegraphics[width=8.50cm]{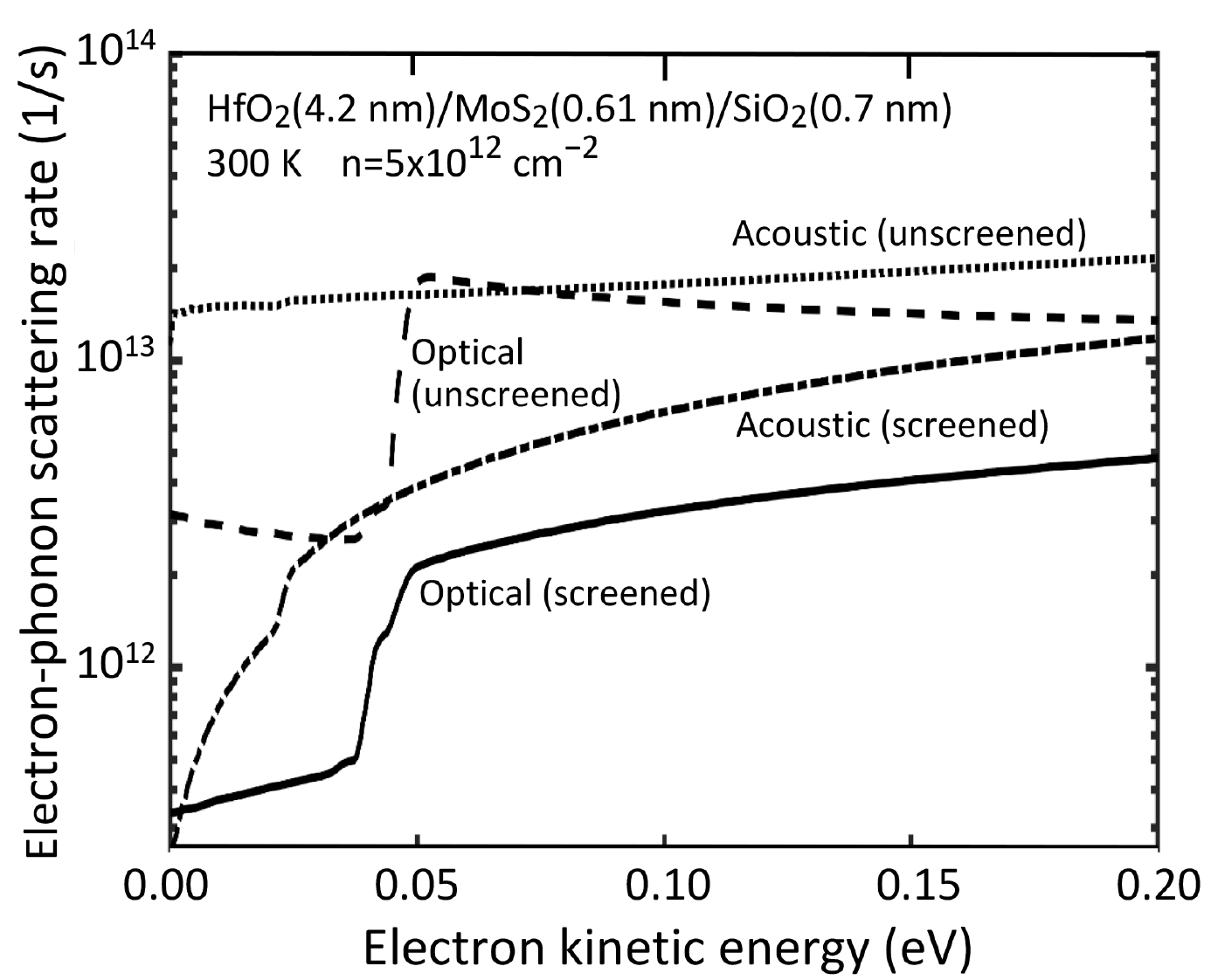}
\caption{Electron-phonon scattering rate in MoS$_2$ as a function of electron kinetic energy for the SiO$_{2}$/MoS$_{2}$/HfO$_{2}$ system accounting for 
         the screening effects of the free carriers (assuming a sheet density of $5 \times 10^{12}$ cm$^{-2}$)
         and of the surrounding dielectrics while ignoring IPP scattering. For comparison, the unscreened `intrinsic' electron-phonon 
         scattering rates of MoS$_{2}$ are also shown. All rates have been averaged over equi-energy surfaces. The physical thickness of each layer
         is shown in parentheses in the legend.}
\label{f:screen_scat}
\end{figure}
\vspace*{-0.25cm}
\subsection{Dielectric screening due to the insulators and free carriers}

Before considering the effects of the all-important IPP scattering, we discuss first how the `intrinsic' carrier mobility in the TMD monolayers 
(that is, limited only by scattering with the bulk TMD phonons) in the double-gate configuration is affected by the dielectric screening due to the 
free carriers in the TMD and the SiO$_{2}$ substrate/bottom-gate-insulator and various gate insulators, ignoring for now IPP scattering.
The gate insulators we have considered range from the low-$\kappa$ SiO$_{2}$ to the high-$\kappa$ ZrO$_{2}$. 

In Fig.~\ref{f:screen_scat} we compare the scattering rates for the unscreened electron/bulk phonons interaction in free-standing MoS$_{2}$ with the screened
rates for the same MoS$_{2}$ layer but embedded in a HfO$_{2}$/MoS$_{2}$/SiO$_{2}$ double-gate structure. The screened rates are more than one order of magnitude
smaller than the unscreened rates, this difference being particularly large at low electron kinetic energies, since dielectric screening is most effective
and long wavelengths and low-energy electrons can transfer only a small (crystal) momentum. 
\begin{figure}[tb]
\includegraphics[width=8.750cm]{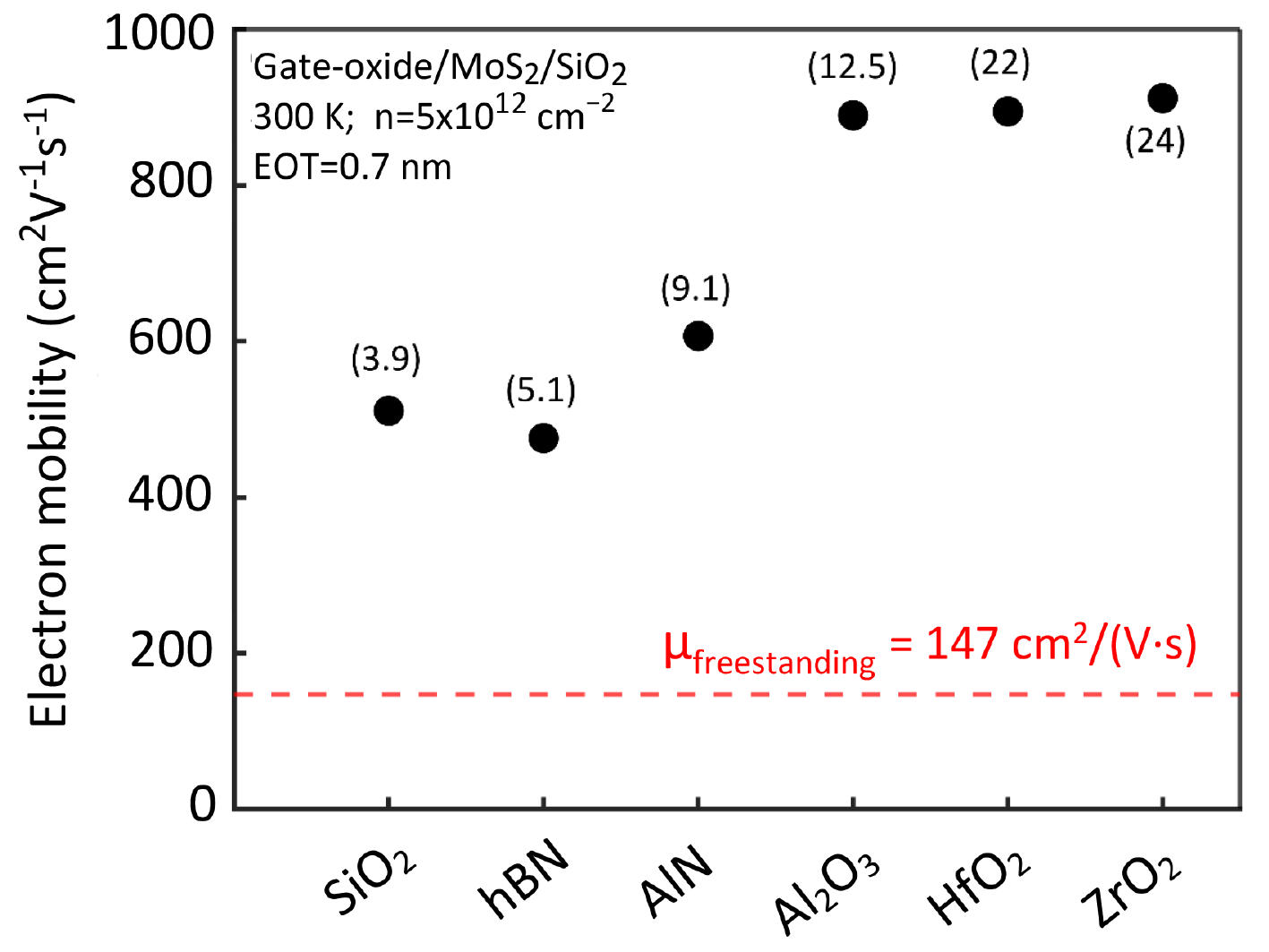}
\caption{Calculated 300~K electron mobility in the TMD channel of a gate-insulator/MoS$_{2}$/SiO$_{2}$ structure including only the effect of 
         free-carrier and dielectric screening by the dielectrics (ordered with increasing static dielectric constant -- shown in parentheses -- 
         from left to right), ignoring IPP scattering. The electron sheet-density in the TMD layer has been assumed to be $5 \times 10^{12}$ cm$^{-2}$.} 
\label{f:screen}
\end{figure}

Such a reduction of the scattering rates results, obviously, in a much higher electron mobility in a layer embedded in double-gate structure than in the
unscreened free-standing configuration, as shown in Fig.~\ref{f:screen}. For reference, the electron mobility of a free-standing MoS$_2$ monolayer 
without such `screening by the environment', is 147~cm$^{2}$/(V.s)~\cite{gaddemane2021limitations}. As expected, the mobility increases with
increasing dielectric constant of the substrate. Screening by a HfO$_{2}$ and ZrO$_{2}$ gate insulator, 
the highest-$\kappa$ dielectrics we consider, increases the phonon-limited electron mobility of MoS$_{2}$ by almost one order of magnitude, 
($>$ 900 cm$^{2}$/(V.s)). Even in the case of a SiO$_{2}$ gate oxide, one can see some modest gains, the electron mobility 
being enhanced to about 500~cm$^{2}$/(V.s). The beneficial effect caused by the presence of high-$\kappa$ dielectrics has been observed experimentally~\cite{Radisavljevic_2011,Radisavljevic_2013,Huo_2018} and theoretically explained both in the case of the impurity-limited electron
mobility in MoS$_2$~\cite{Ong_2013a} as well as of the phonon-limited mobility~\cite{Huo_2018}, although in this latter case the observed 
`phonon quenching' and improvement of the mobility was attributed to HfO$_2$-induced tensile strain. 

\subsection{Electron-IPP scattering}
\label{sec:IPP-scattering}

\begin{figure}[tb]
\includegraphics[width=8.5cm]{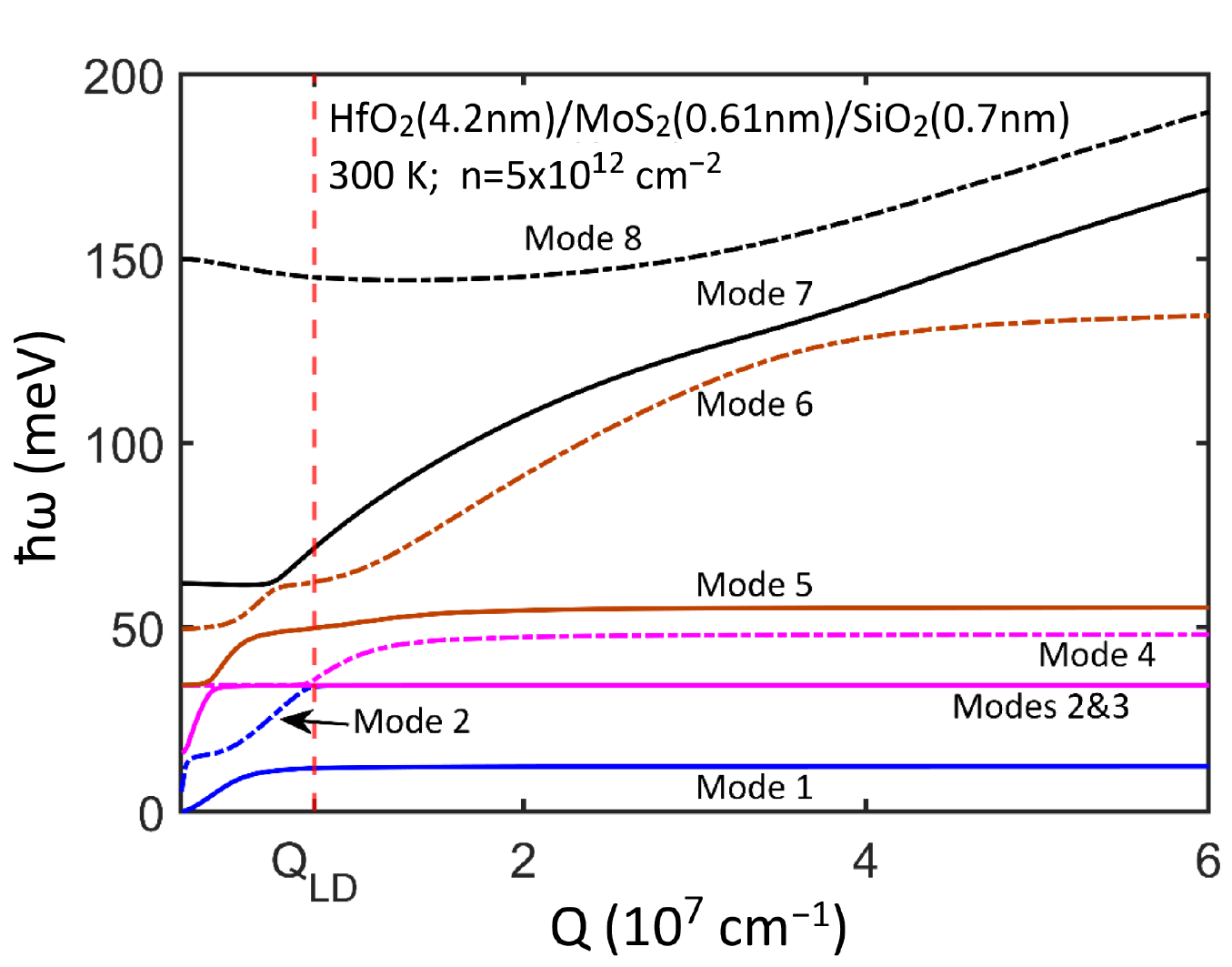}
\caption{Dispersion of the fully hybridized IPPs for the SiO$_2$/MoS$_2$/HfO$_2$ structure. The wave vector 
         $Q_{\rm LD}$ ($\approx 7.8 \times 10^{6}$ cm$^{-1}$ at the sheet density shown in the legend) indicates the wavelength-cutoff
         beyond which the plasmons are Landau-damped. For $Q > Q_{\rm LD}$, only statically-screened SO phonons survive.}
\label{fig:disp_HfO2_MoS2}
\end{figure}

As already anticipated, IPP scattering negates completely the benefits we have just discussed. In order to understand the nature of this scattering process, 
we first need to consider the dispersion of these hybrid modes. This is shown in Fig.~\ref{fig:disp_HfO2_MoS2} for the HfO$_{2}$/MoS$_{2}$/SiO$_{2}$ double-gate
structure. These results have been obtained assuming an electron sheet-density of $5 \times 10^{12}$ cm$^{-2}$ and room and temperature (300~K) 
by solving the secular equation, Eq.~\ref{eqn:seculardg} for each Q. This is an 8$^{\rm th}$-degree algebraic equation that we have solved 
using a combination of the Newton's method, of the bisection technique, and by finding the eigenvalues of the `companion matrix'
associated to the polynomial appearing at the left-hand-side of Eq.~(\ref{eqn:seculardg})~\cite{Horn_1985}. We must stress again that, as explained in
sec~\ref{sec:Landau}, we consider the full dispersion relation shown in this figure only for $Q \leq Q_{\rm LD}$.

In the case considered in Fig.~\ref{fig:disp_HfO2_MoS2}, we can see clearly the full hybridization of the modes, especially around $Q \sim$ 1/nm. 
Only at large $Q$ do the modes become almost pure plasmon- or phonon-like excitations. In particular, the label `mode 1' refers to the lowest-energy 
IPP that is mainly a 2D plasmon at small $Q$ and becomes mainly the low-frequency TO phonon of HfO$_{2}$ at large $Q$. Similarly, at large $Q$, modes 2-5
originate, respectively, mainly from the from the two antisymmetric interface MoS$_{2}$ (mostly ZO-like) interface phonons~\cite{Zhang_2020}, 
from the low-frequency SiO$_2$ and from the high-frequency HfO$_2$ phonons. 
In the same range of large $Q$, mode 6 originates mainly from the high-frequency SiO$_2$ TO phonon, whereas, finally, modes 7 and 8 represent mainly 
the 2D plasmons at the bottom (MoS$_2$/SiO$_2$) and top (MoS$_2$/HfO$_2$) interfaces. 

\begin{figure}[tb]
\includegraphics[width=8.75cm]{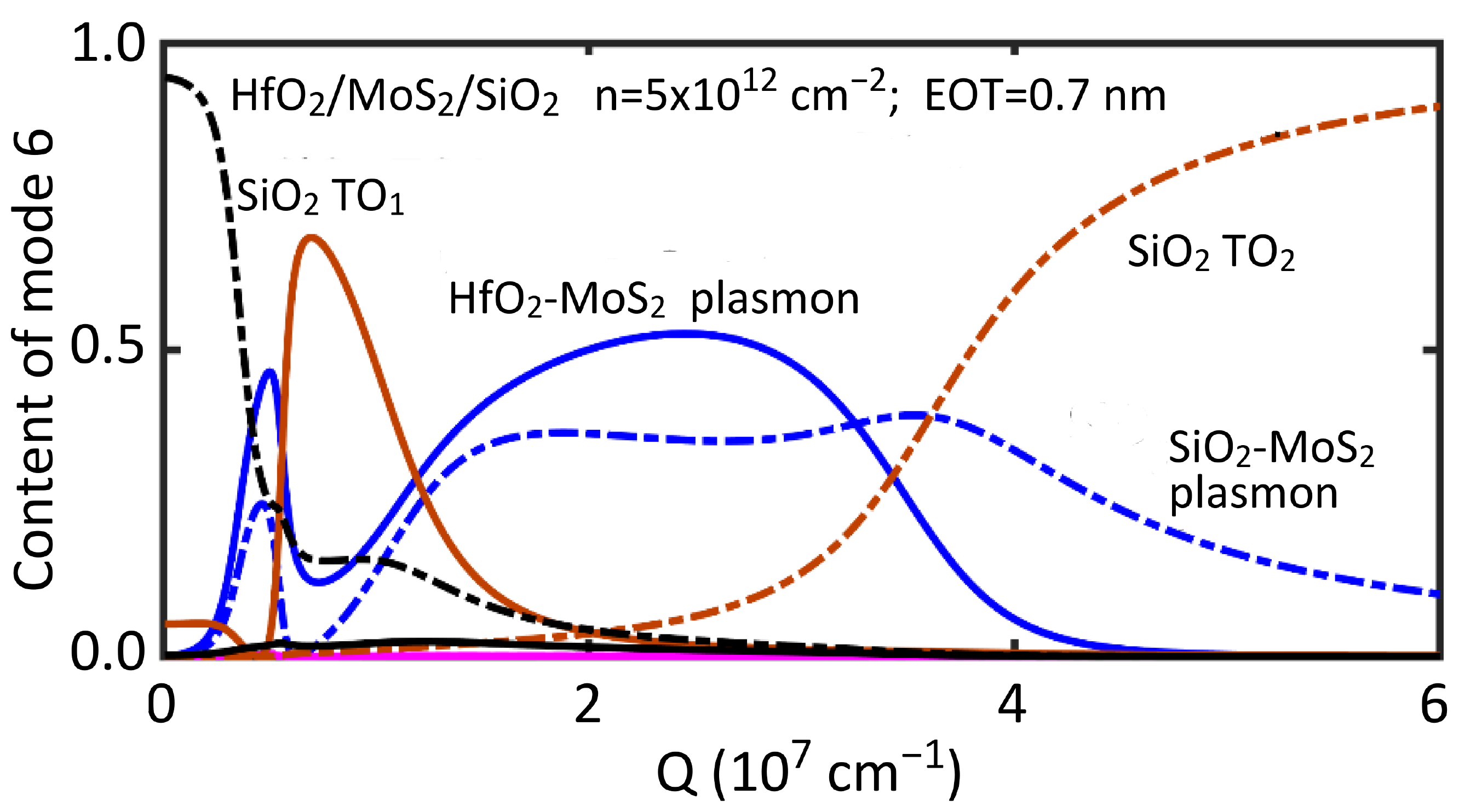}
\caption{Phonon and plasmon content of the fully hybridized IPP labeled `mode 6' in Fig.~\ref{fig:disp_HfO2_MoS2}.}
\label{fig:content_mode6_HfO2}
\end{figure}
\begin{figure}[tb]
\includegraphics[width=7.950cm]{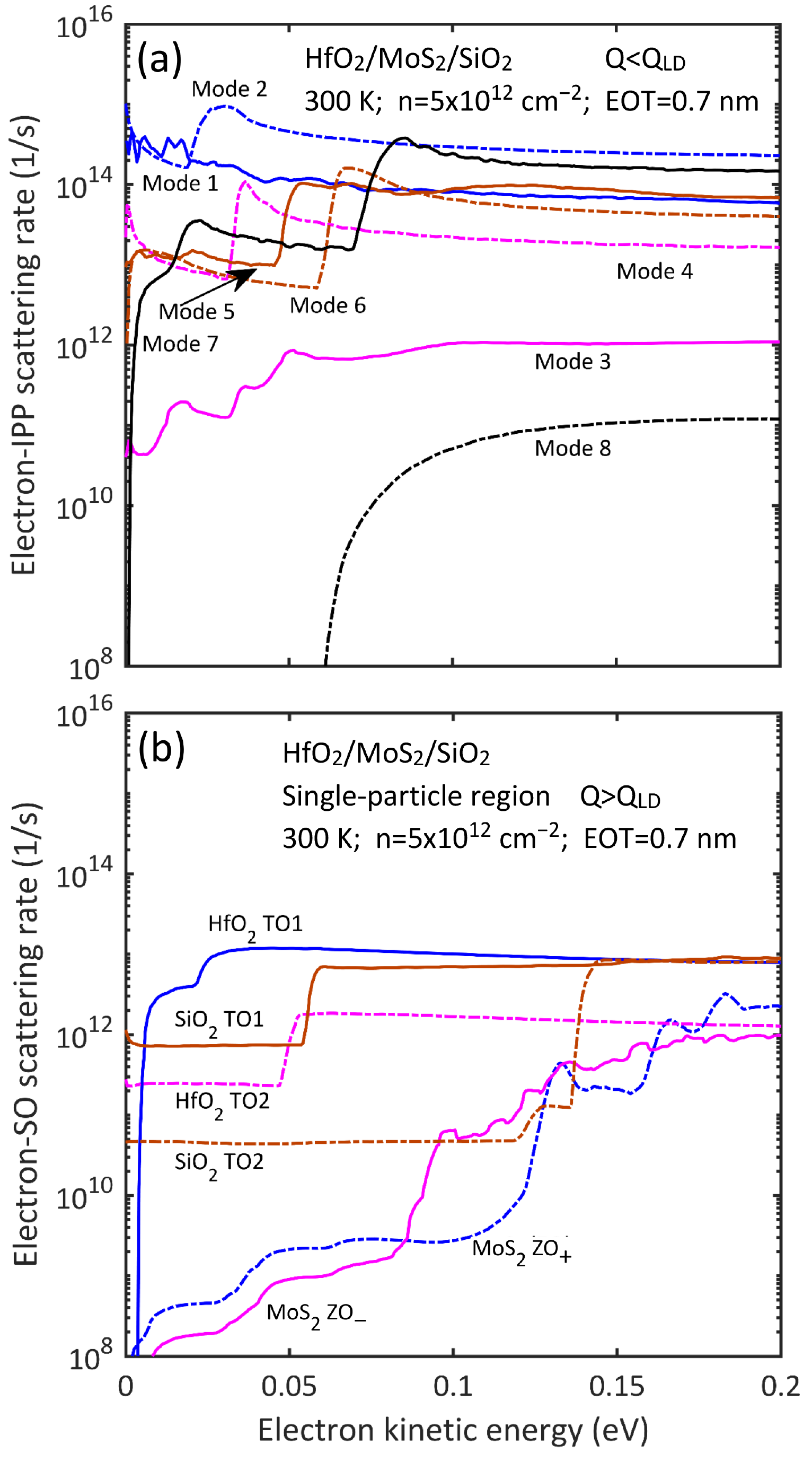}
\caption{Electron-IPP scattering rates {\it{vs.}} electron kinetic energy in the SiO$_2$/MoS$_2$/HfO$_2$ stack calculated for a carrier density 
         of $5 \times 10^{12}$ electrons cm$_{-2}$ and at 300~K. The anisotropic rates -- functions of the electron wave vector ${\mathbf {k}}$ --
         have been averaged over equi-energy surfaces. The rates shown in the top frame, {\bf (a)}, have been computed considering only final 
         states such that the wave vector transfer $Q$ is smaller than the
         Landau-damping cutoff $Q_{\rm LD} \approx 7.8 \times 10^{6}$ cm$^{-1}$.
         They represent interactions only with the long-wavelength fully hybridized IPPs. The rates for scattering 
         only with the short-wavelength SO modes (optical phonons hybridized among themselves but decoupled from the Landau-damped plasmons) are shown in 
         the bottom frame, ${\bf (b)}$. A comparison with Fig.~\ref{fig:disp_HfO2_MoS2} permits the identification of the eight modes
         shown in frame {\bf (a)}.}
\label{fig:IPPrates_HfO2}
\end{figure}\
\begin{figure}[tb]
\includegraphics[width=8.0cm]{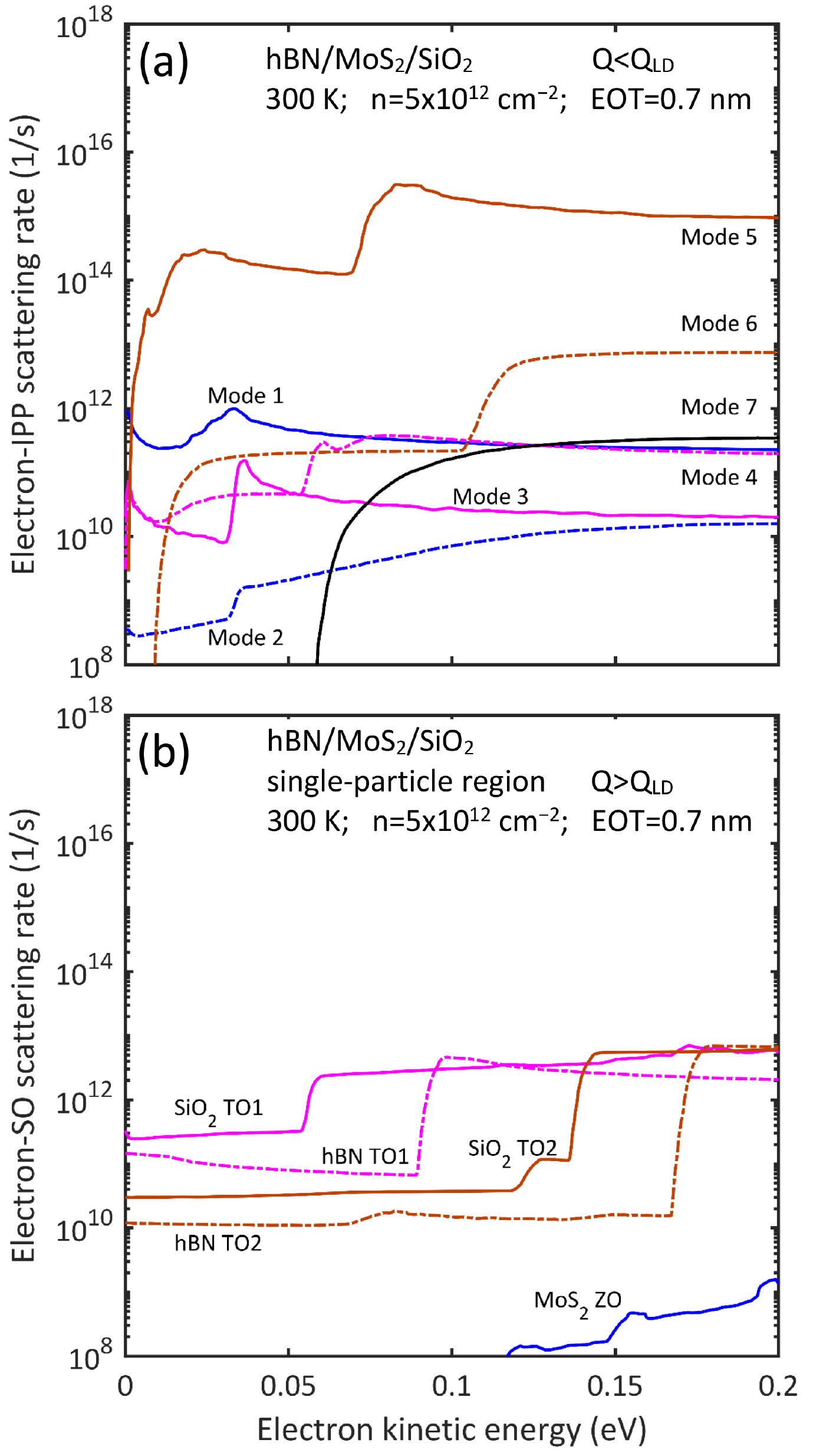}
\caption{As in Fig.~\ref{fig:IPPrates_HfO2}, but for the SiO$_2$/MoS$_2$/hBN stack.}
\label{fig:IPPrates_hBN}
\end{figure}
\begin{figure}[tb]
\includegraphics[width=8.5cm]{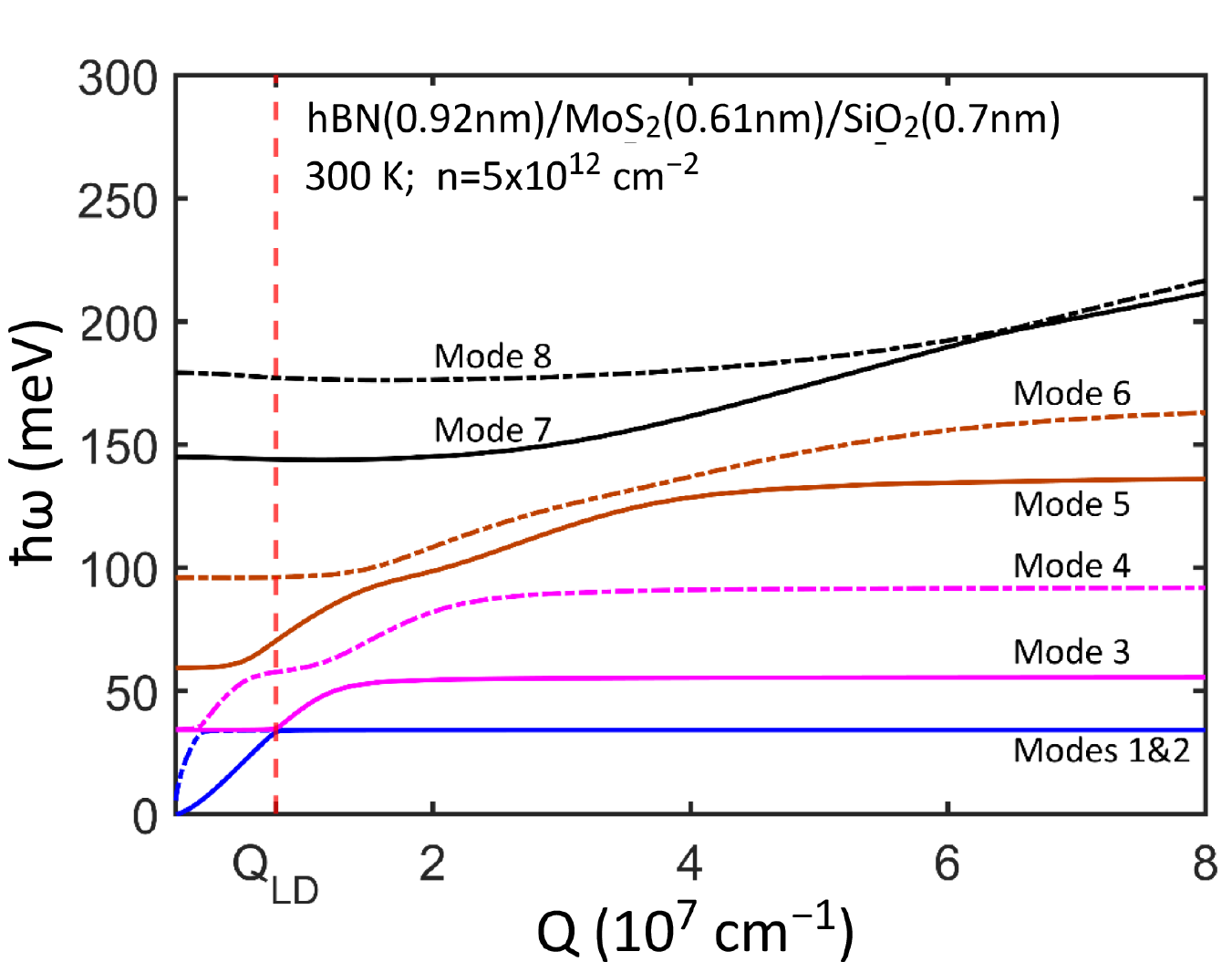}
\caption{Dispersion of the fully hybridized IPPs for the SiO$_2$/MoS$_2$/hBN structure. }
\label{fig:disp_hBN_MoS2}
\end{figure}\

An example of the complicated hybridization of these modes is illustrated in Fig.~\ref{fig:content_mode6_HfO2}: The phonon, 
$\Phi^{(\alpha)}(\omega^{(5)}_{Q})$, and plasmon content of the decoupled mode 6 shown in Fig.~\ref{fig:disp_HfO2_MoS2}
is plotted as a function of wave vector $Q$.  At small $Q$, this excitation is mainly a low-energy SiO$_2$ phonon (TO1); as its wavelength decreases, it
hybridizes with the high-frequency HfO$_2$ phonon, with the MoS$_2$ optical phonon, with both MoS$_2$ interface plasmons (at the lower and upper interfaces),
becoming mainly a high-frequency SiO$_2$ phonon as $Q \rightarrow \infty$. For $Q \approx 10^{6}$/cm, this mode hybridizes 
significantly with up to 5 different bare modes. As shown below, this is the long-wavelength, non Landau-damped region in which 
scattering with these modes controls electronic transport; therefore, this hybridization is an essential physical ingredient, as already shown in
Refs.~\cite{Fischetti_2001,Ong_2012,Ong_2013} for the case of supported and gated graphene.
    
For $Q > Q_{\rm LD}$ ($\approx 7.8 \times 10^{6}$ cm$^{-1}$ in MoS$_2$ for an electron sheet density of $5 \times 10^{12}$~cm$^{-2})$, 
the 8 hybrid branches shown in Fig.~\ref{fig:disp_HfO2_MoS2} are replaced by six surface-phonon-like (SO) branches 
obtained by solving the secular equation, Eq.~(\ref{eqn:seculardg}), but ignoring the plasma response of the 2DEG by replacing 
$(1-\omega_{\rm P}(Q)^{2}/\omega^{2}$) in Eq.~(\ref{eq:eps2D1}) 
with the static Thomas-Fermi expression, $1+Q_{\rm TF}^{2}/Q^{2}$ (where $Q_{\rm TF}$ is the screening wave vector given by Eqs.~(\ref{eq:QTFnd}) 
and (\ref{eq:QTFdeg}).

Very important is Fig.~\ref{fig:IPPrates_HfO2} in which we show the electron-IPP scattering rates separately for interactions with the 8 long-wavelength, 
fully hybridized IPPs (Fig.~\ref{fig:IPPrates_HfO2}(a)) and with the interface optical phonons (SOs) hybridized among themselves but decoupled from 
the Landau-damped 2D plasmons (Fig.~\ref{fig:IPPrates_HfO2}(b)). These rates have been obtained integrating Eq.~(\ref{eq:rate5}) only for values
of the transfer wave vector $Q$ smaller (Fig.~\ref{fig:IPPrates_HfO2}(a)) or larger (Fig.~\ref{fig:IPPrates_HfO2}(b)) 
than the Landau-damping cut-off $Q_{\rm LD}$. As for the results of Fig.~\ref{fig:disp_HfO2_MoS2}, the calculations have been performed assuming 
an electron sheet-density of $5 \times 10^{12}$~cm$^{-2}$ and room temperature (300~K). The results shown in this figure show that, despite the fact that 
Landau damping prevents plasmons to exists over a large region of momentum-space, nevertheless the $\sim 1/Q^{2}$-behavior of the squared electron-IPP matrix
elements causes scattering to be dominated processes involving ${\bf Q}$-vectors of magnitude small enough to render the role of phonon/plasmon coupling 
very important. Of course, such a small-$Q$ scattering process results in momentum-relaxation rates that are not as strong as the scattering rates. 
Yet, coupling to plasmons is an effect that cannot be neglected. This also shows that, at the electron sheet-density of interest, considering the 
long-wavelength limit, $Q < Q_{\rm LD} \approx 10^{6}-10^{7}$~cm$^{-1}$ $\ll 2\pi/h \approx 10^{8}$~cm$^{-1}$, appears to be a sensible approximation. 

Both in the long-wavelength regime as well as in the single-particle region,  
transport is controlled by the HfO$_2$ phonons (modes 1 and 2) and, to some extent, by the low-frequency phonon of the SiO$_2$ substrate.
Note that the scattering rates labeled `ZO$_{\pm)}$' in frame {\bf {(b)}} refer to the interaction with the weak non-ZO-component of  
SO excitations at the top and bottom TMD/insulator interfaces, excitations that are mainly TMD optical flexural modes.  
A comparison of Fig.~\ref{fig:IPPrates_hBN} with Fig.~\ref{fig:IPPrates_HfO2} shows how a low-$\kappa$ gate-insulator, like hBN, induces negligible IPP
scattering. Also in this case, electron-IPP scattering is dominated by the long-wavelength processes. However, the strongest scattering channel is due
to the interaction with the SiO$_2$ phonons (see also Fig.~\ref{fig:disp_hBN_MoS2}), especially with the low-frequency mode at small electron energies.    

We shall now discuss how IPP scattering affects the carrier mobility {\emph{i)}} with different gate insulators, {\emph{ii)}} as a function of their
thickness, {\emph{iii)}} of the dielectric constant assumed for the TMD monolayer, {\emph{iv)}} of the density of the 2DEG, and, finally, 
{\emph{v)}} on temperature.\\
 
\begin{figure*}[t]
\includegraphics[width=17.50cm]{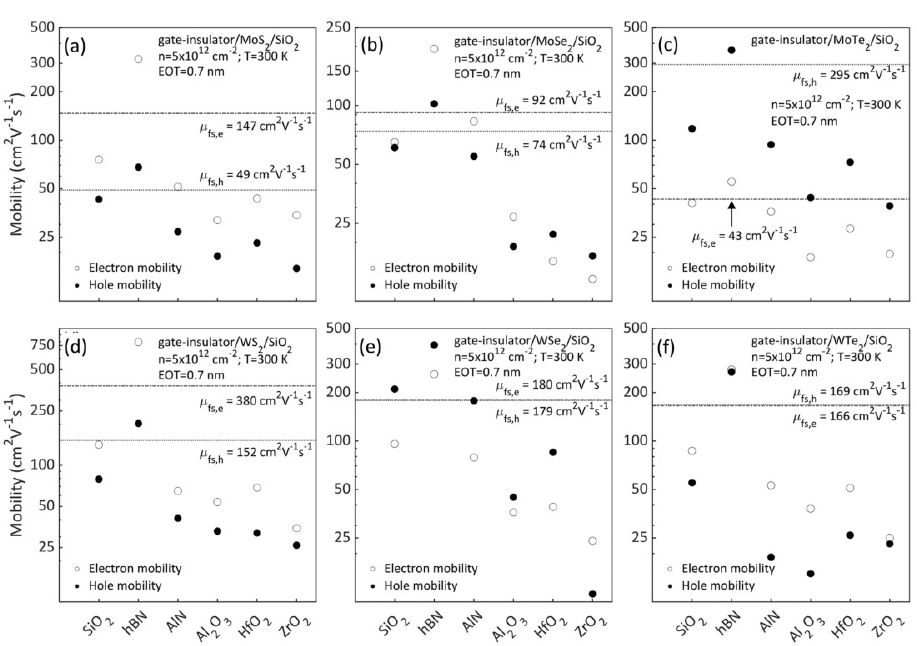}
\caption{{\bf{(a)}}: Calculated 300 K electron (open symbols) and hole mobility (solid symbols) in the TMD channel of a gate-insulator/MoS$_{2}$/SiO$_{2}$ 
         structure with different gate insulators (ordered with increasing static dielectric constant – shown in parentheses -- from left to right). 
         Note that the mobility generally decreases as the static dielectric constant of the gate insulator increases, but phonon resonances cause ‘dips’ 
         in the cases of Al$_{2}$O$_{3}$ and, to a smaller extent, ZrO$_{2}$. Moreover, the absence of low-energy phonons in hBN results in a mobility larger 
         than in free-standing MoS$_{2}$ (shown by the horizontal lines), thanks to the beneficial effect of dielectric screening of the bulk phonons by 
         high-$\kappa$ insulators. {\bf{(b)}}: The same, but for the gate-insulator/MoSe$_{2}$/SiO$_{2}$ structure and {\bf{(c)}}, for the 
         gate-insulator/MoTe$_{2}$/SiO$_{2}$ structure. One can notice a similar behavior. Finally, frames {\bf{(d)}}, {\bf{(e)}} and {\bf{(f)}} shows the
         same information for the gate-insulator/WS$_{2}$/SiO$_{2}$, gate-insulator/WSe$_{2}$/SiO$_{2}$, and gate-insulator/WTe$_{2}$/SiO$_{2}$
         systems, respectively. In all cases, the carrier sheet density in the TMD layers has been assumed to be $5 \times 10^{12}$ cm$^{-2}$.}         
\label{f:topoxide}
\end{figure*}\
\begin{table}[tb] 
\caption{Calculated 300 K electron mobility (in cm$^2$V$^{-1}$s$^{-1}$) in TMD monolayers in a double-gate stack, assuming SiO$_{2}$ as bottom insulator 
         and different gate insulators. Both insulators are assumed to have an equivalent oxide thickness (EOT) of 0.7~nm. The results have been obtained
         assuming a TMD electron sheet-density of $5 \times 10^{12}$ cm$^{-2}$.}
\vspace*{0.5cm}
\begin{tabular}{lllllll}
\hline
\hline
System                        &     MoS$_{2}$ &  MoSe$_{2}$ &  MoTe$_{2}$  &   WS$_{2}$ &   WSe$_{2}$  &   WTe$_{2}$$^{ a}$ \\
\hline
Freestanding                  &      147      &   92        &       43     &    380     &      180     &   166      \\
SiO$_{2}$/TMD/SiO$_{2}$       &      76       &   65        &       41     &    141     &       96     &    87      \\
hBN/TMD/SiO$_{2}$             &      319      &   195       &       55     &    795     &      260     &   278      \\
AlN/TMD/SiO$_{2}$             &      51       &   83        &       36     &    65      &       79     &    53      \\
Al$_{2}$O$_{3}$/TMD/SiO$_{2}$ &      32       &   27        &       19     &    54      &       36     &    38      \\
HfO$_{2}$/TMD/SiO$_{2}$       &      43       &   16        &       28     &    69      &       39     &    51      \\
ZrO$_{2}$/TMD/SiO$_{2}$       &      34       &   13        &       20     &    35      &       24     &    25      \\
\hline
\hline
\end{tabular}\\
{\small \flushleft \hspace*{-6.5cm} $^{a}$ See note~\cite{bnote3}}\\
\label{tab:electron_mobility}
\end{table}
\begin{table}[tb] 
\caption{Calculated 300 K hole mobility (in cm$^{2}$V$^{-1}$s$^{-1}$) in the TMD channel of several double-gate stacks, as in Table~\ref{tab:electron_mobility}.}
\vspace*{0.5cm}
\begin{tabular}{llllllll}
\hline
\hline
System                        &     MoS$_{2}$ &  MoSe$_{2}$ &  MoTe$_{2}$  &   WS$_{2}$ &   WSe$_{2}$   &   WTe$_{2}$$^{\ a}$   \\
\hline
Freestanding                  &      49       &    74       &    295       &     152    &     179       &       169             \\
SiO$_{2}$/TMD/SiO$_{2}$       &      43       &    61       &    118       &      79    &     210       &        55             \\
hBN/TMD/SiO$_{2}$             &      68       &   102       &    363       &     202    &     396       &       268             \\
AlN/TMD/SiO$_{2}$             &      27       &    55       &     94       &      41    &     178       &        19             \\
Al$_{2}$O$_{3}$/TMD/SiO$_{2}$ &      19       &    19       &     44       &      33    &      45       &        15             \\
HfO$_{2}$/TMD/SiO$_{2}$       &      23       &    22       &     73       &      32    &      85       &        26             \\
ZrO$_{2}$/TMD/SiO$_{2}$       &      16       &    17       &     39       &      26    &      11       &        23             \\
\hline
\hline
\end{tabular}\\
{\small \flushleft \hspace*{-6.5cm}$^{a}$ See note~\cite{bnote3}}\\
\label{tab:hole_mobility}
\end{table}
\noindent\hspace*{-0.5cm}{\emph{{i) Dependence on the gate insulator}}.
In Fig.~\ref{f:topoxide} we show the carrier mobility calculated in TMD monolayers assuming various top insulators, listed in order of increasing dielectric constant,
from SiO$_{2}$ to ZrO$_{2}$, for the double-gate structure we have considered, assuming SiO$_2$ as substrate. Also in this case, we have assumed
a carrier sheet-density of $5 \times 10^{12}$~cm$^{-2}$ and room temperature. For convenience, we list the calculated values also in
Tables~\ref{tab:electron_mobility} and \ref{tab:hole_mobility} 

Among the various supported and gated TMDs considered, the best electron mobility, 795~cm$^{2}$/(V.s), is obtained for the SiO$_{2}$/WS$_{2}$/hBN system. 
An excellent hole mobility, 396~cm$^{2}$/(V.s) is also predicted for the SiO$_{2}$/WSe$_{2}$/hBN stack. These results confirm the general trend 
observed previously for MoS$_{2}$~\cite{van2020effects}: The carrier mobility decreases almost monotonically with increasing dielectric constant
of the gate insulator, with two exceptions: {\bf{1}}. The beneficial effects of dielectric screening of the ‘out-of-plane’ field lines are seen for 
hBN, since electron-IPP scattering is weak in this case, thanks to its relatively low ionic polarization and the high phonon frequencies resulting from 
the light weight of the B and N ions. A similar result has been obtained by Ma and Jena~\cite{Ma_2014}.
{\bf{2}}. On the contrary, resonance effects among the optical phonons of the substrate, of the TMD layer, 
and of the top insulator result in a low carrier mobility when AlN and/or Al$_{2}$O$_{3}$ are taken as gate insulators. 
In particular, looking at Fig. \ref{f:topoxide}(a), one notices that the
calculated 300~K electron and hole mobilities in SiO$_2$-supported monolayer MoS$_{2}$ decrease as the gate insulator is changed from hBN to HfO$_{2}$, but it 
reaches a particularly high value for hBN and a very low value for Al$_{2}$O$_{3}$. A similar behavior, although to a different extent, can be seen also 
for most of the other cases shown in Fig. \ref{f:topoxide}. In all cases, hBN exhibits the best mobility, even higher 
than what we have calculated in free-standing monolayers, as explained above. Obviously, this comes at the price of the relatively low dielectric 
constant -- that defies the very reason behind the use of high-$\kappa$ insulators, as dictated by scaling laws -- and a relatively small potential barrier 
at the channel-insulator interface -- that would lead to a large gate-leakage current if hBN were to be used as the sole gate dielectric.\\   

{\emph{The role of resonances}}. 
The low mobility seen for Al$_{2}$O$_{3}$ and ZrO$_{2}$ in almost all cases is caused by fact that the low-frequency optical phonon 
of Al$_{2}$O$_{3}$ (or high-frequency ZrO$_{2}$ TO phonon) oscillates at a frequency that is very close to the frequency of the low-energy SiO$_{2}$ TO 
phonons and of the hybridized MoS$_{2}$ ZO phonon. In these cases, these modes couple 
very strongly and, as a result of this resonance, the large interface polarization-charge strengthens the IPP scattering potential. 
A similar effect has been already predicted theoretically for the Si/SiO$_{2}$/poly-Si-gate system. When the poly-Si density yields a plasma frequency 
close to the TO-frequency of SiO$_{2}$, the resonance of these two modes causes a low carrier mobility 
(Refs.~\cite{kotlyar2004inversion,Shah_2007,suleiman2012effects}). 
Obviously, one concern about the accuracy of these results stems from the consideration that such an effect is extremely sensitive to the phonon frequencies chosen
for the calculations (approximated from the broad peaks seen by Fourier-transform infrared spectroscopy, FTIR~\cite{Fischetti_2001}, for example) 
and observed in dielectrics used in experiments (whose morphology and composition may or may not be stoichiometric and amorphous/polycrystalline).
Small changes of the phonon frequency may result in significantly different values for the mobility. \\

\noindent{\emph{ii) Dependence on insulator thickness}}.
Figure~\ref{f:thickness} shows the dependence of the carrier mobility -- in the same double-gate systems -- on the thickness of the gate insulator. 
As a general trend, we observe that the mobility decreases as the thickness of the insulator increases. This is due to the screening of the IPP scattering-potential due
to the proximity of an ideal metal gate. In the HfO$_{2}$/MoS$_{2}$/SiO$_{2}$ system, we see a 20\% increase of the mobility when the 
equivalent-SiO$_2$ oxide thickness (EOT) decreases from 1.2 to 0.7~nm. A similar trend is seen in the HfO$_{2}$/WS$_{2}$/SiO$_{2}$ stack, 
although the effect is stronger. This is due to the fact that the electron mobility in free-standing WS$_{2}$ is larger than in MoS$_{2}$, so that the 
mobility is much more sensitive to any changes of the strength of IPP scattering.\\
\begin{figure}[tb]
\includegraphics[width=8.25cm]{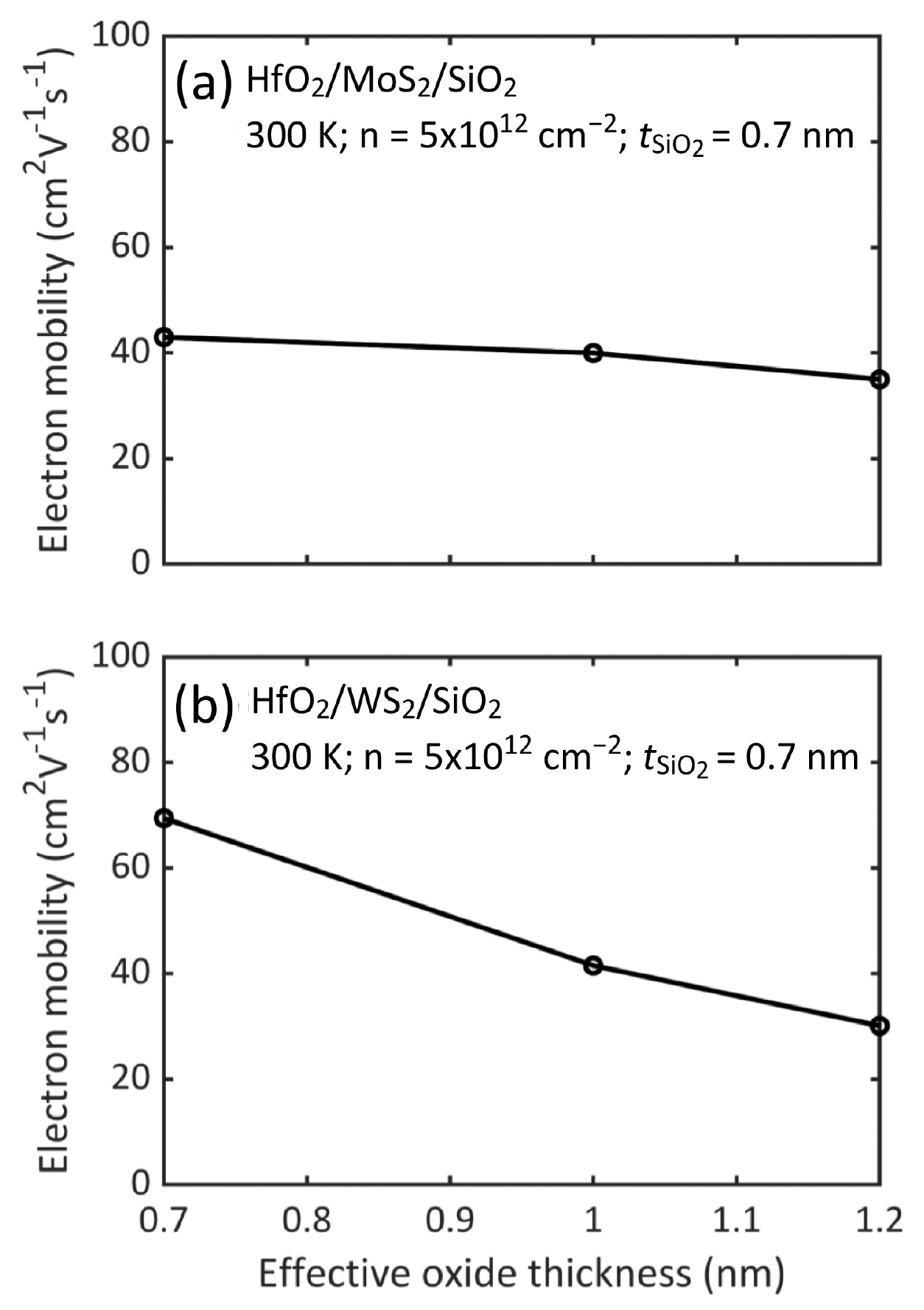}
\caption{{\bf{(a)}}: Calculated 300 K electron mobility in the TMD channel of a HfO$_{2}$/MoS$_{2}$/SiO$_{2}$ structure with a different thickness of the
          gate insulator; {\bf{(b)}}: The same, but for the HfO$_{2}$/WS$_{2}$/SiO$_{2}$ structure. A carrier sheet density of $5 \times 10^{12}$ cm$^{-2}$
          has been assumed in these calculations. The symbols represent calculated data, the lines are only a guide to the eye.}
  \label{f:thickness}
\end{figure}\

\noindent{\emph{iii) Dependence on the dielectric constant of the TMD monolayer}}.
All values given so far for the carrier mobility have been obtained using the `effective' static and optical dielectric constants for the TMD layers. 
However, given the anisotropy of the dielectric constant of the TMD layers and the various approximations used in the literature and here, it is interesting 
to see how the mobility varies as this parameter is changed. For MoS$_2$, for example, the static and the high-frequency out-of-plane dielectric constants are 
6.2$\epsilon_0$ and 6.1$\epsilon_0$ (Ref.~\cite{laturia2018dielectric}), respectively; the effective and geometric-averaged static(optical) values are  
9.8(9.69)$\epsilon_0$, and 11.42(11.31)$\epsilon_0$. As a trend, the electron mobility increases with increasing TMD dielectric constant.
In particular, for the HfO$_{2}$/MoS$_{2}$/SiO$_{2}$ stack, the mobility obtained using the out-plane dielectric constant is about 38~cm$^2$V$^{-1}$s$^{-1}$. 
This value increases by approximately 20\%, to 43~cm$^2$V$^{-1}$s$^{-1}$ when using the effective dielectric constant, and by almost 60\% 
(to 58~cm$^2$V$^{-1}$s$^{-1}$) when using the geometric average for the dielectric constant.
Such a trend, also reported by Hauber and Fahy~\cite{Hauber_2017}, is just due to the expected role of dielectric screening. \\

\noindent{\emph{iv) Dependence on carrier density}}.
In Fig.~\ref{f:density}, looking at the variation of mobility with varying free-electron density, we observe only a marginal improvement
when the density increase from 10$^{\rm 11}$ cm$^{-2}$ to 10$^{\rm 13}$ cm$^{-2}$. Since the free carriers are confined in the 2D plane of the
layer, they cannot screen efficiently the long-wavelength electron-phonon scattering processes. Additionally, degeneracy effects also affect
the mobility: As the Fermi level increases, Pauli blocking prevents scattering to already occupied states, an effect that results in a higher mobility. 
However, this effect is partially compensated by the fact that, as the Fermi level increases, scattering to satellite valleys become effective, resulting in a reduction of the electron mobility. In the particular case of MoS$_{2}$, these degeneracy effects cancel each other almost exactly and do not affect the mobility appreciably. This behavior can be expected also in other 2D TMDs. The weak dependence on the carrier sheet density shown in Fig.~\ref{f:density} has been observed experimentally by Pang {\it {et al.}} in FETs with thick ($>$2~nm) TMD channels.~\cite{Pang_2021}.\\

\begin{figure}[tb]
\includegraphics[width=8.25cm]{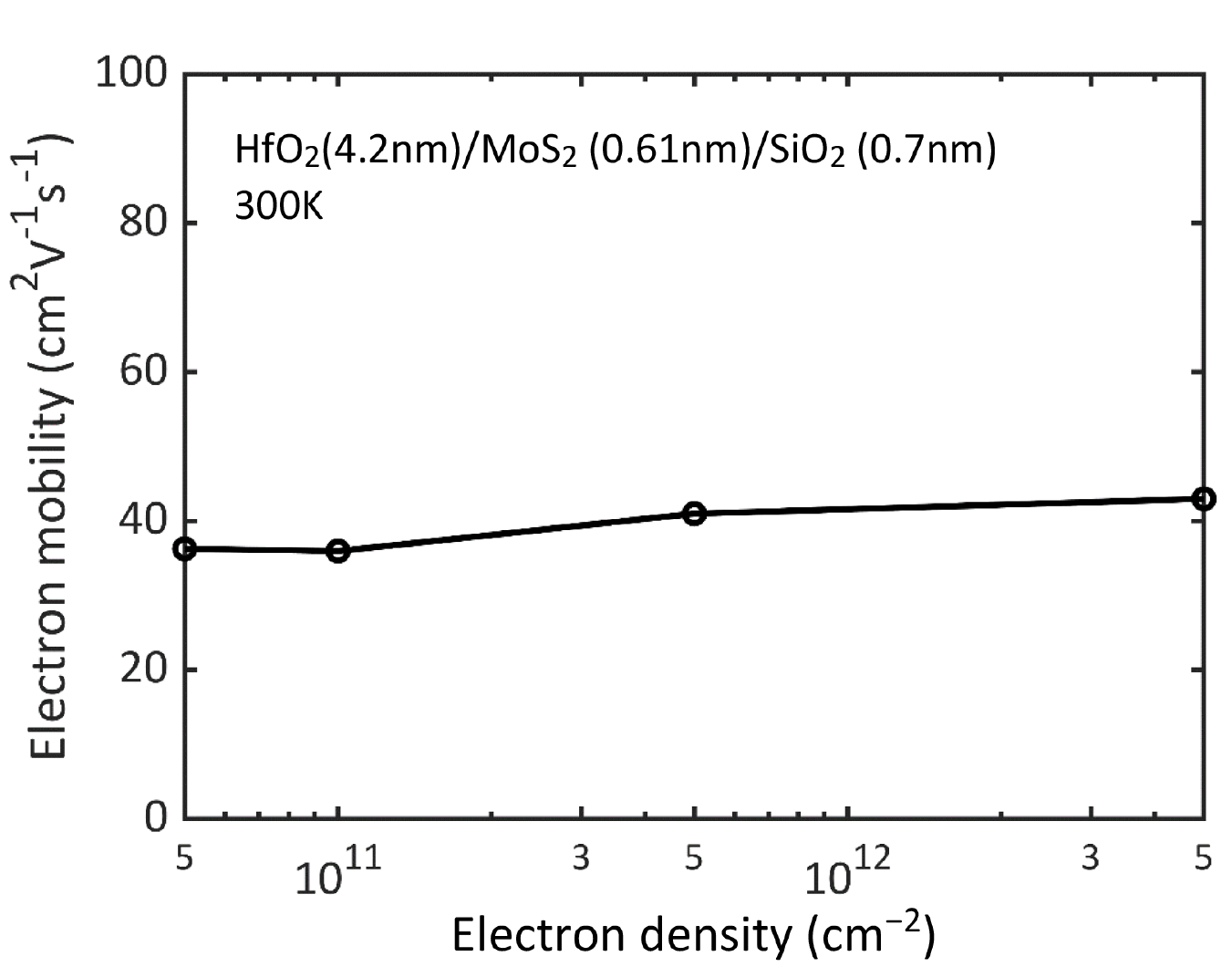}
\caption{Calculated 300~K electron mobility in the TMD channel of a HfO$_{2}$/MoS$_{2}$/SiO$_{2}$ structure as a function of the TMD free-carrier density.
         An electron sheet density of $5 \times 10^{12}$ cm$^{-2}$ has been assumed.
         The symbols represent calculated data, the lines are only a guide to the eye.}
  \label{f:density}
\end{figure}\
\begin{figure}[tb]
\includegraphics[width=8.0cm]{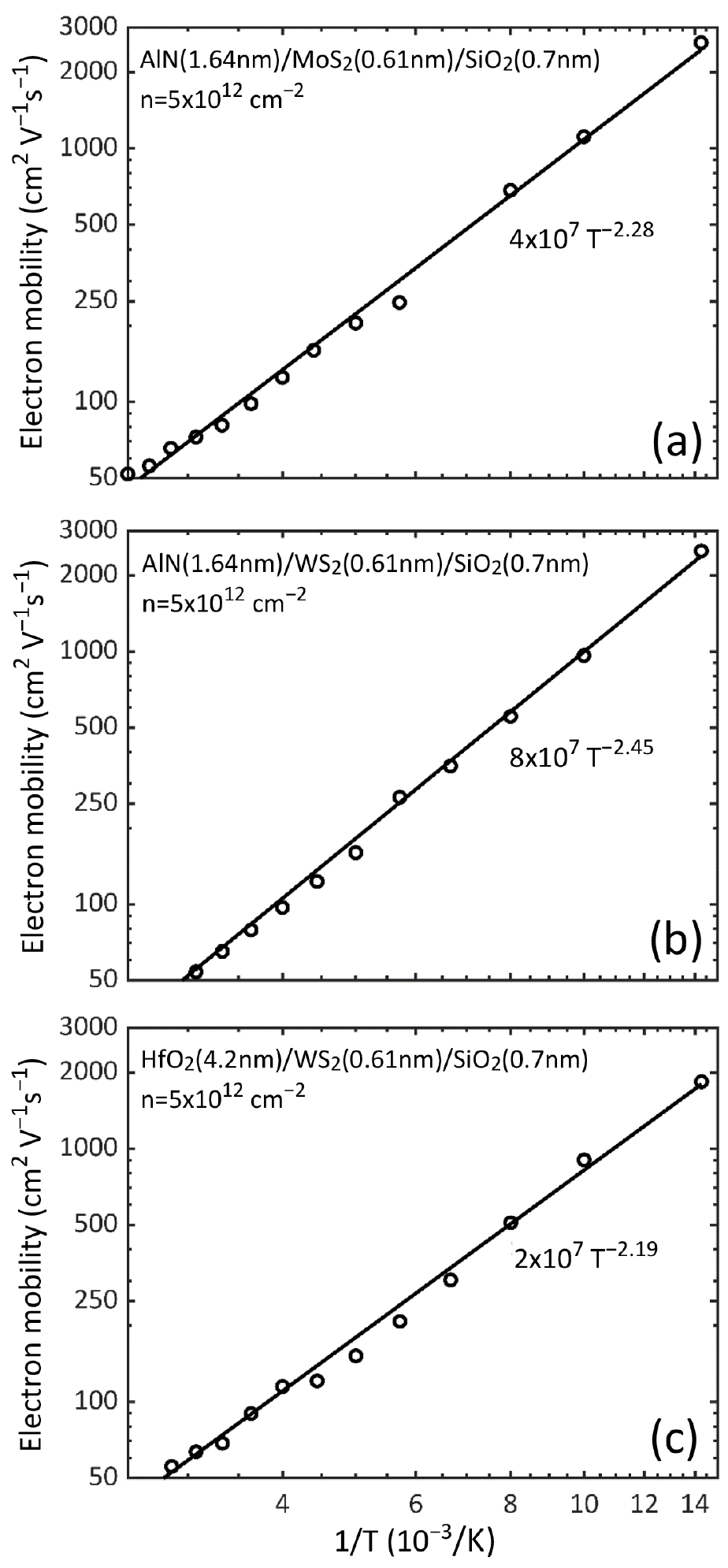}
\caption{Calculated electron mobility in the TMD channel of the (a) AlN/MoS$_{2}$/SiO$_{2}$, (b) AlN/WS$_{2}$/SiO$_{2}$, and 
         (c) HfO$_{2}$/WS$_{2}$/SiO$_{2}$ structures as a function of lattice temperature.
         As in previous figures, the EOT of the insulators has been assumed to be 0.7~nm and the electron sheet density $5 \times 10^{12}$ cm$^{-2}$.}
\label{f:temperature}
\end{figure}\
\begin{figure}[tb]
\includegraphics[width=8.0cm]{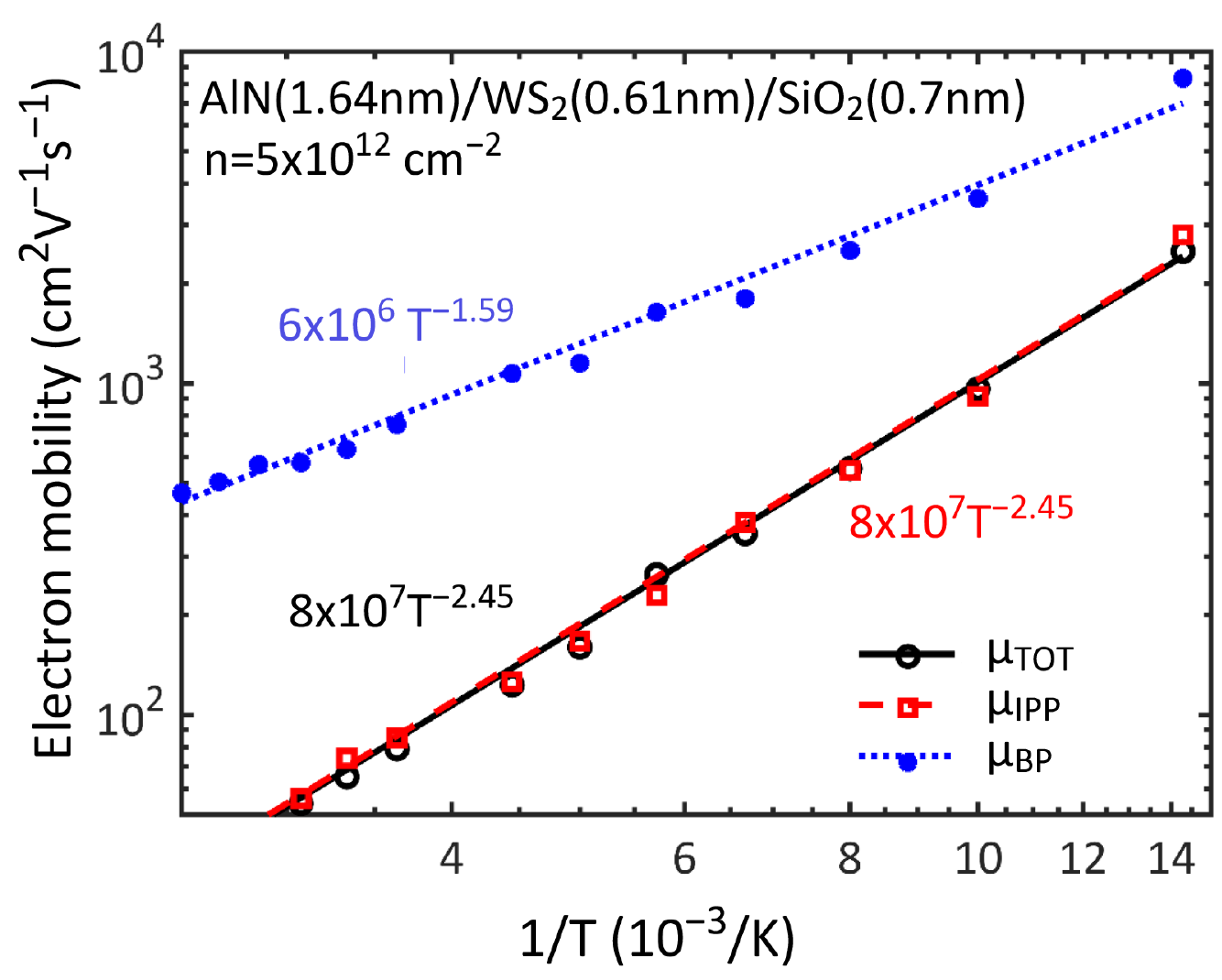}
\caption{Calculated bulk-phonon-limited, $\mu_{\rm BP}$, IPP-limited, $\mu_{\rm IPP}$, and total electron mobility, $\mu_{\rm TOT}$, 
         in the TMD channel of an AlN/WS$_{2}$/SiO$_{2}$ stack, plotted {\it {vs}}. the lattice temperature.
         As in previous figures, the EOT of the insulators has been assumed to be 0.7~nm and the electron sheet density $5 \times 10^{12}$ cm$^{-2}$.
          }
\label{fig:Tsplit}
\end{figure}\
\noindent{\emph{v) Dependence on lattice temperature}}.
The dependence of the electron mobility on temperature is shown in Fig.~\ref{f:temperature} for the the AlN/MoS$_{2}$/SiO$_{2}$, 
AlN/WS$_{2}$/SiO$_{2}$, and HfO$_{2}$/WS$_{2}$/SiO$_{2}$ double-gate system. A power law (shown by the solid black lines) yields the best correlation coefficient.
However, it is difficult to assign to this fit a clear physical meaning, since several phonons with different frequencies contribute to the relaxation
of the electron velocity. Comparing frames (a) and (b), the slightly larger
frequency of the TMD ZO phonon seems to yield a slightly larger slope. The same observation can be made comparing frames (b) and (c): The softer
optical phonons of HfO$_2$ result in a smaller slope (and a lower mobility).  

We plot in Fig.~\ref{fig:Tsplit} the temperature dependence of the electron mobility in the TMD monolayer in the AlN/WS$_{2}$/SiO$_{2}$ stack, showing 
separately the mobility limited only by the TMD `bulk' phonons, $\mu_{rm BP}$ and the mobility limited only by scattering with the IPPs, $\mu_{\rm IPP}$,
Note the weaker dependence on temperature of the former, since $\mu_{\rm BP}$ is determined by the interaction with both low-energy ($\lesssim$ 20~meV)
acoustic phonons and optical phonons with energy in the range of 40-to-50~meV. On the contrary, the IPP-limited mobility shows a stronger 
dependence on temperature, since it is controlled mainly by the interaction with the phonon component of hybrid modes originating from the higher-energy 
($\sim$ 80~meV) AlN optical phonons. Scattering with the hybrid IPPs controls the mobility over the entire range of temperatures we have considered, although
its relative importance decreases at low temperatures. It is interesting to note that the use of Matthiessen's rule to estimate the total mobility as
$\mu_{\rm TOT} = [1/\mu_{\rm BP}+1/\mu_{\rm IPP}]^{-1}$ results in a value that is underestimated by about 7\% at 400~K, by about 17\% at 70~K.  

Power laws for the temperature dependence of the electron mobility in TMD monolayers have been observed experimentally at carrier densities in the range of the
low $10^{11}$s to the high $10^{12}$s cm$^{-2}$, with data scattered significantly, not a surprising fact given the many process-dependent effects and deviations
from ideality that unavoidably affect the measurements. To mention just a small set of experimental results reported in the literature, Wang and 
coworkers~\cite{Wang_2021} found a power-law dependence, $\mu \sim T^{-\gamma}$, with an exponent $\gamma$ of about 1.73 for WS$_{2}$ monolayers encapsulated 
by hBN. The mobility they report ranges from 800~cm$^{2}$/(V$\cdot$s) at 100~K to around 120~cm$^{2}$/(V$\cdot$s) at 300~K, values that, although not too
dissimilar from the results shown in Figs.~\ref{f:temperature} and \ref{fig:Tsplit} in the presence of the high-$\kappa$ AlN insulator, indicate the strong 
effect of additional scattering mechanisms, such as Coulomb scattering with charges defects. Indeed, for the same system, Xu {\it {et al.}}~\cite{Xu_2016} 
have measured a steeper slope, with an exponent of about 1.9 -- and of about 2.3 for similarly hBN-encapsulated MoS$_{2}$ monolayers -- with a significantly
higher mobility ranging from 3,000~cm$^{2}$/(V$\cdot$s) at 100~K to around 300~cm$^{2}$/(V$\cdot$s) at 200~K. An exponent of about 0.73, with mobilities in the
range of 45 to 100~cm$^{2}$/(V$\cdot$s) between 250 and 100~K, has been observed by Ovchinnikov {\it et al.}}~\cite{Ovchinnikov_2014} for back-gated MoS$_{2}$
monolayers on SiO$_{2}$. Also for back-gated MoS$_2$ supported by SiO$_2$, Smithe {\it{et al.}}~\cite{Smithe_2018} have measured an exponent of about 1.24 with 
a mobility decreasing from 200~cm$^{2}$/(V$\cdot$s) at 80~K to $\sim$35~cm$^{2}$/(V$\cdot$s) at 300~K, whereas an exponent of about 1.66, albeit with a very 
low mobility, has been reported in Ref.~\cite{Alharbi_2016} for WS$_{2}$ monolayers, also on an SiO$_{2}$ substrate. Finally, Huo and coworkers~\cite{Huo_2018}
have measured an exponent of about 0.7 for the HfO$_{2}$/MoS$_{2}$/SiO$_{2}$ gate stack. The measured mobility varies from $\sim$80~cm$^{2}$/(V$\cdot$s) at 
100~K to about 50~cm$^{2}$/(V$\cdot$s) at 300~K, low values probably due to the combined effect of IPP and additional scattering processes. 
With the exception of Ref.~\cite{Wang_2021}, that reports results from four-point probe measurements, and of Ref.~\cite{Xu_2016}, that reports 
both the Hall and the field-effect mobility -- all experiments provide the field-effect mobility. In all cases, Coulomb scattering with defects, and even possible
thermally-activated transport below 100~K~\cite{Huo_2018} and variable-range hopping among traps in the low-temperature regime 
(20-to-250~K)~\cite{Ovchinnikov_2014},
depresses the slope as well as the mobility, even showing saturation of the mobility below 100~K. Therefore it is not surprising to see a weaker dependence than
what we have obtained for ideal monolayers in the absence of scattering with defects and hopping processes. 

\section{Conclusions}
\label{sec:Conclusions}

We have investigated theoretically how, in addition to free-carrier screening, the dielectric environment (namely, the insulating substrate or bottom gate-insulator, the gate insulator, and the presence of metallic gates) affects the low-field charge-transport properties of several TMD monolayers supported (or back-gated)
by SiO$_2$ and with hBN, SiO$_2$, AlN, Al$_2$O$_3$, HfO$_2$, and ZrO$_2$ as gate insulators, thus extending previous
investigations to these realistic (and complicated) double-gate structures.
 
We have shown initially that, rescaling the electron-phonon scattering rates obtained from {\it {ab initio}} calculations for free-standing layers to account
for the static screening effects of the surrounding dielectrics, high-$\kappa$ dielectrics screen these interactions very
efficiently. Although, in principle, this would result in an improved carrier mobility, unfortunately, the presence of these dielectrics also results
in carriers scattering with the hybrid interface optical-phonon/plasmon excitations. 
Accounting for the full hybridization of these excitations also with the 2D TMD plasmons-- an effect that we found is very important, but that is often ignored
in the literature -- we have shown that, as expected, this effect depresses the carrier mobility
significantly below its value in free-standing monolayers, with the mobility increasing with increasing dielectric constant of the insulator(s).
However, we have also found that `phonon resonances' enhance the interface polarization charge and scattering potential, thus depressing the mobility even
more severely. This occurs when two optical phonons of the dielectrics have similar frequencies and it results in an even stronger depression of the mobility.  

Our results also show that the weak ionic polarizability of hBN results in the highest electron and hole mobility, although the relatively low dielectric constant
of this insulator negates the scaling benefits of high-$\kappa$ insulators. They also show that tungsten-based TMDs exhibit a satisfactory carrier mobility. 
We have also found a weak dependence of the mobility on carrier density, that the dielectric screening effect of an ideal metal gate is noticeable in
some cases (especially in TMDs that exhibit a high intrinsic mobility, such as WS$_2$), but not significantly strong. Unfortunately, in most cases
the transport properties of TMDs remain disappointing, because of both the effects we have investigated here as well as, most likely, because of the
intrinsic physical reasons discussed in Ref.~\cite{Cheng_2020}. 
\vspace*{-0.25cm}
\acknowledgments
We are grateful to J\"{o}rg~Appenzeller, Zhihong~Chen, and Eric~Pop for having shared preliminary experimental data and/or constructive discussions 
and to Edward Chen for support and encouragement. We also thank Mathieu Luisier for clarifications regarding Ref.~\cite{Fiore_2022} and Brian~Ridley for many conversations that have clarified the confusion caused by the use of the term `remote phonon scattering' to refer improperly to IPP scattering. 

This work has been funded by the Semiconductor Research Corporation (SRC) nCORE/NEWLIMITS program and, in part,
by the Taiwan Semiconductor Manufacturing Company Ltd. (TSMC).

Portions of this work were presented and published as an extended abstract at the 2022 International Conference on Simulation of Semiconductor Processes and 
Devices\cite{Gopalan_2022}.  

\bibliography{paper.bib}

\end{document}